\documentclass[12pt,preprint]{aastex}

\def\beq{\begin{equation}}
\def\enq{\end{equation}}
\def\bea{\begin{array}}
\def\ena{\end{array}}

\begin{document}

\title{Detectability of Long GRB Afterglows from Very High Redshifts}
\author{ L. J. Gou$^1$, P. M\'esz\'aros$^{1,2,3}$, T. Abel$^1$, \& B. Zhang$^1$}
\affil{$^1$Dept. Astronomy \& Astrophysics, 525 Davey Lab, Pennsylvania
State University, University Park, PA 16802}
\affil{$^2$Dept. of Physics, 104 Davey Lab, Pennsylvania
State University, University Park, PA 16802}
\affil{$^3$Institute for Advanced Study, Princeton, NJ 08540}


\begin{abstract}

Gamma-ray bursts are promising tools for tracing the formation of
high redshift stars, including the first generation. At very high
redshifts the reverse shock emission lasts longer in the observer
frame, and its importance for detection and analysis purposes
relative to the forward shock increases. We consider two different
models for the GRB environment, based on current ideas about the
redshift dependence of gas properties in galaxies and primordial
star formation. We calculate the observed flux as a function of
the redshift and observer time for typical GRB afterglows, taking
into account intergalactic photoionization and Lyman-$\alpha$
absorption opacity as well as extinction by the Milky Way Galaxy.
The fluxes in the X-ray and near IR bands  are compared with the
sensitivity of different detectors such as Chandra, XMM, Swift XRT
and JWST.  Using standard assumptions, we find that Chandra, XMM
and Swift XRT can potentially detect GRBs in the X-ray band out to
very high redshifts $z\gtrsim$ 30. In the K and M
bands, the JWST and ground-based telescopes are potentially able
to detect GRBs even one day after the trigger out to $z\sim$ 16
and 33, if present. While the X-ray band is insensitive to the
external density and to reverse shocks, the near IR bands provide
a sensitive tool for diagnosing both the environment and the
reverse shock component.

\end{abstract}



\section{Introduction}

Gamma-ray bursts are thought to be associated with the formation
of massive stars (van Paradijs, Kouveliotou, \& Wijers 2000). The
evidence for this has been mainly in the class of long bursts, of
$\gamma$-ray durations in excess of 2 seconds, making up
two-thirds of the GRB population, which are the only ones so far
for which X-ray, optical, IR and radio afterglows, as well as
redshifts, have been measured. The strongest evidence yet comes
from the recently confirmed association of long GRBs with
core-collapse supernovae (Stanek et al. 2003; Hjorth et al. 2003;
Uemura et al. 2003; Price et al. 2003). Short bursts, of durations
less than 2 seconds, even if produced e.g. by neutron star
mergers, would similarly be associated with massive star
formation, and one expects the rate of occurrence of GRBs with
redshift to follow closely the massive star formation rate. In
currently favored $\Lambda$CDM cosmologies, star formation should
start at redshifts higher than those where proto-galaxies and
massive black holes at their centers develop (Miralda-Escud\'e
2003). Thus, GRBs could trace the pre-galactic star formation era
preceding quasars.

Recent cosmic microwave background anisotropy data collected by
WMAP reveal that the first objects in the Universe should be
formed around $z \approx 18$ (Bennett et al.  2003). This is
consistent with the theoretical modeling of the first star
formation (Abel et al. 1998, 2000, 2002; Bromm et al. 1999). There
is also indirect observational evidence for high-z GRBs. E.g.,
empirical relations have been found between the GRB luminosities
and other measured quantities, such as the variability of the
gamma-ray light curves (Fenimore \& Ramirez-Ruiz 2000) and
spectral lags (Norris et al. 2000). By extrapolating these
empirical laws to a larger burst sample (e.g. the BATSE data), it
is found that many BATSE bursts would be expected to have $z>6$
(Fenimore \& Ramirez-Ruiz 2000).

The discovery of the highest redshift quasars, such as the current
record holder at $z=6.43$ (Fan et al. 2003), grows increasingly difficult
because the quasar formation rate drops rapidly at higher
redshifts, peaking between redshift 2 and 3. Very few galaxies can
be seen above $z>6$, which is also consistent with the upper limit
for the redshift of galaxy formation $z_{gal} \le 9$ based on
theoretical analysis (e.g. Padmanadhan 2001). Although young
galaxies may exist at very high redshifts, they are likely to be
too faint to obtain good spectra (Haiman \& Loeb 1997). On the
other hand, the extreme brightness of GRBs during their first day
or so make them the most luminous astrophysical objects in the
Universe. Thus, GRBs appear to be promising tools to explore the
very high redshift Universe (Miralda-Escud\'e, 1998).

The natural question which needs to be quantified is the degree of
detectability of GRBs with current or future detectors, if they
occur at much higher redshifts than those currently sampled.  Lamb
\& Reichart (2000) used specific templates such as GRB 970228
observed at one day to estimate the highest redshifts at which
such bursts could be observed using Swift. Ciardi \& Loeb (2000)
calculated the flux evolution with redshift of common GRBs and
discussed the flux change with redshift at several epochs in the
infrared bands. These papers considered only forward shock
radiation as known before 2000 and some effects of the galactic
mean density evolution but did not consider the primeval
star-formation environment.

In this paper we have calculated the flux evolution of typical
GRBs based on current knowledge about GRB physics in a more
realistic way. Among the refinements introduced are: (1) The
contribution from reverse shocks is considered as a crucial
element. This should be very important for the early afterglow in
the rest-frame (which at high redshifts gets dilated to longer
observed times). Therefore, we expect that at higher redshifts the
possibility of observing the reverse shock is much increased. (2)
We have taken up to date GRB parameters, e.g. incorporating new
estimates of the typical magnetic equipartition parameter
$\epsilon_B$ about one order or more magnitude smaller than the
electron parameter $\epsilon_e$ in the forward shock, and a possibly
higher $\epsilon_B$ in the reverse
shock. This has a significant effect on the GRB evolution. (3) We
consider GRB external densities motivated both by views on the
typical protogalaxy density evolution with redshift, and by views
on the conditions around the first stars to form in the universe
in the pre-galactic era. (4) We consider both the Lyman-$\alpha$
and photoionization absorption as well as our own galactic
extinction. (5) We compare the expected fluxes  in the X-ray and
near IR bands to the sensitivity of various detectors such as
Chandra, XMM, Swift XRT and JWST.

In \S\S 2.1 and 2.2 we outline the basic forward and reverse shock
flux calculations, the details of which are given in an appendix.
We discuss the GRB density environment in \S 2.3, and the
intergalactic and galactic absorption effects are estimated in \S
2.4.  In \S\S 3.1 and 3.2 we discuss the optical/IR and X-ray flux
dependence on redshift, respectively, at various observer times,
including the dependence on external density. We compare these to
the Swift XRT, Chandra, XMM and JWST sensitivities for the
detection of GRBs at different redshifts. We summarize the
numerical results and discuss the implications in \S 4.


\section{Afterglow characteristics}

\subsection{Forward Shock}

We assume that the shock-accelerated electrons have a power-law
distribution of Lorentz factors $\gamma_e$ with a minimum Lorentz
factor $\gamma_m$: $N(\gamma_e)d\gamma_e \propto
\gamma_e^{-p}d\gamma_e,\gamma_e \ge \gamma_m$. We also define a
critical Lorentz factor $\gamma_c$ above which the electrons cool
radiatively on a time shorter than the expansion time scale
(M\'esz\'aros, Rees, \& Wijers 1998). This leads to the standard
(forward shock) broken power law spectrum of GRBs (Sari, Piran, \&
Narayan 1998). In the fast-cooling regime, when
$\gamma_m>\gamma_c$, all the electrons cool rapidly down to a
Lorentz factor $\approx \gamma_c$ and the observed flux at
frequency $\nu$ is

\begin{equation}
 F_{\nu}= F_{\nu,m,f}\left\{ \begin{array}{l@{\quad \quad}l}
              (\nu/\nu_{c,f})^{1/3} & \nu < \nu_{c,f} \cr
              (\nu/\nu_{c,f})^{-1/2}  &  \nu_{c,f} \le \nu < \nu_{m,f} \cr
              (\nu_{m,f}/\nu_{c,f})^{-1/2}(\nu/\nu_{m,f})^{-p/2} & \nu_{m,f} \le \nu
          \end{array} \right.
\label{eq:fluxforfast}
\end{equation}
Hereafter the subscripts `f' and `r' indicate forward and reverse
shock, respectively.

In the slow-cooling regime, when $\gamma_c> \gamma_m$, only electrons
with $\gamma_e >\gamma_c$  cool efficiently, and the observed flux is
\begin{equation}
 F_{\nu}= F_{\nu,m,f}\left\{ \begin{array}{l@{\quad \quad}l}
              (\nu/\nu_{m,f})^{1/3} & \nu<\nu_{m,f} \cr
              (\nu/\nu_{m,f})^{-(p-1)/2}  &  \nu_{m,f} \le \nu < \nu_{c,f} \cr
              (\nu_{c,f}/\nu_{m,f})^{-(p-1)/2}(\nu/\nu_{c,f})^{-p/2} & \nu_{c,f} \le \nu
          \end{array} \right.
\label{eq:fluxforslow}
\end{equation}
where $F_{\nu,m,f}$ is the observed peak flux at the observed
frequency $\nu=\min(\nu_{c,f}, \nu_{m,f})$, while $\nu_{m,f}$ and
$\nu_{c,f}$ are
the observed frequencies corresponding to $\gamma_m$ and
$\gamma_c$, respectively. Synchrotron self-absorption can also
cause an additional break at very low frequencies, typically about
$\le 5 \ {\rm GHz}$, in the radio range. Since here we focus on
the IR and X-ray ranges, we will not consider this low-frequency
regime in our calculations. For a fully adiabatic shock, the
evolution of the typical frequency and peak flux are given by Sari
et al. (1998):

\begin{eqnarray}
\nu_{c,f}&=&2.5 \times10^{12}
\epsilon_{B,f}^{-3/2}E_{52}^{-1/2}n^{-1} t_5^{-1/2}\times
(1+z)^{-1/2}\ {\rm Hz},  \\
 \label{eq:nuc}
 \nu_{m,f}&=&4.6 \times10^{14} \epsilon_{B,f}^{1/2}
\epsilon_e^2E_{52}^{1/2}t_5^{-3/2} \times (1+z)^{1/2}\ {\rm Hz}, \\
\label{eq:num}
 F_{\nu,m,f}&=&1.1 \times 10^5
\epsilon_{B,f}^{1/2}E_{52} n^{1/2}D_{28}(z)^{-2}\times (1+z)\ {\rm \mu Jy~}.
\label{eq:fnum}
\end{eqnarray}
Hereafter the quantities without the subscript `s' are in the observer
frame, and the quantities with the subscript `s' are for the observer
in the local frame of the source, which is connected with the
observer frame quantities with a certain power of $(1+z)$. The source
is assumed at a
luminosity distance $D_L(z)=10^{28}D_{28}(z)$ cm, and $\epsilon_B$
and $\epsilon_e$ are the fraction of the shock energy converted
into energy of magnetic fields and accelerated electrons,
respectively. The time is taken in units of $t=10^5 t_5$ s
($\simeq$ 1 day), $E_{52}=E/10^{52}\ {\rm ergs}$ is the isotropic
equivalent energy of the GRBs, and $n$ is the particle density in
units of ${\rm cm^{-3}}$ in the ambient medium around the GRB.

\subsection{Reverse Shock}

As GRBs are measured at increasingly larger redshifts, a given
constant observer time corresponds to increasingly shorter source
frame times. This is favorable for observing at very high
redshifts the evolution of phenomena which happen only in the
earliest stages of the GRB, such as the reverse shock emission. So
far, the reverse shock emission has been observed in only three GRBs in the
optical band: GRB 990123 (Akerlof et al. 1999), GRB 021004 (Fox et
al. 2003a) and GRB 021211 (Fox et al. 2003b; Li et al. 2003). At
these early epochs, the reverse shock emission makes a significant
contribution to the overall flux of the GRB afterglow. A
description of the reverse shock spectrum is however more
complicated than that of the forward shock. It depends on two
factors: (1) whether one is in the thick shell or thin shell case,
and (2) the ratio of the crossing time of the reverse shock across
the shell to the observing time. We consider a relativistic shell
with an isotropic equivalent energy $E$ and initial Lorenz factor
$\eta\equiv L_\gamma/{\dot M} c^2$ expanding into a homogeneous
interstellar medium of particle number density $n$. In the local
frame, we can define a deceleration timescale when the accumulated
ISM mass is $1/\eta$ of the ejecta mass, $t_{dec,s}=[(3E/4\pi
\eta^2 n m_p c^2)^{1/3}/2 \eta^2 c]$, which is the conventional
deceleration timescale. A critical initial Lorenz factor $\eta_c$
can be defined by the condition that the deceleration time
$t_{dec,s}$ equals the intrinsic (i.e. central engine dominated)
duration $T_s$ of the gamma-ray burst, which is $\eta_c\simeq
228.6 E_{52}^{1/8}n^{-1/8}T_{s,1}^{-3/8}$. The thick shell case
occurs when the duration $T_s>t_{dec,s}$, and the thin shell case
occurs when $T_s<t_{dec,s}$. The time taken by the reverse shock to
cross the shell is defined as $t_{\times,s}=\max(t_{dec,s}, T_s)$.
In the observer frame, $t_\times=t_{\times,s}(1+z)$. For
observation times $t<t_\times$, the reverse shock emission
spectrum qualitatively resembles the forward shock spectrum.
However, for $t>t_\times$, we take the approximation that
there is no reverse shock emission above $\nu_{c,r}$, since all
electrons have cooled below that energy. Thus the reverse shock
spectrum at $t>t_\times$ is, in the fast cooling case
\begin{equation}
 F_{\nu}= F_{\nu,m,r}\left\{ \begin{array}{l@{\quad \quad}l}
              (\nu/\nu_{c,r})^{1/3} & \nu < \nu_{c,r} \cr
              0  &  \nu_{c,r}\leq \nu < \nu_{m,r} \cr
              0  & \nu_{m,r} \leq \nu
          \end{array} \right.
\label{eq:fluxrevfast}
\end{equation}
In the slow cooling case the reverse shock spectrum is
\begin{equation}
 F_{\nu}= F_{\nu,m,r}\left\{ \begin{array}{l@{\quad \quad}l}
              (\nu/\nu_{m,r})^{1/3} & \nu < \nu_{m,r} \cr
              (\nu/\nu_{m,r})^{-(p-1)/2}  &  \nu_{m,r} \le \nu < \nu_{c,r} \cr
              0                       & \nu_{c,r} \le \nu
          \end{array} \right.
\label{eq:fluxrevslow}
\end{equation}
where $\nu_{c,r},\nu_{m,r}$ and $F_{\nu,m,r}$ refer here to the
reverse shock cooling frequency, typical frequency and peak flux,
respectively.  Since these quantities are usually different in the
reverse and in the forward shocks, and have different functional
forms and time evolution dependences in the thick and thin shell
cases, the specific shock and shell cases will be differentiated
in the treatment below.

Kobayashi (2000) has given expressions of cooling frequency,
typical frequency of electrons and peak flux in the reverse shock.
Motivated by recent observations of prompt flashes, a set of
relations linking the cooling frequency, typical frequency and
peak flux in the reverse and forward shocks at the crossing time
was proposed by Kobayashi \& Zhang (2003) and  Zhang, Kobayashi,
\& M\'{e}sz\'{a}ros (2003). The flux calculated with these two
different sets of formulae are consistent within a 10\% error
range. Here we use the relationships as discussed in the last two
quoted references,

\begin{equation}
\frac{\nu_{m,r}(t_{\times})}{\nu_{m,f}(t_{\times})}=(\gamma_{\times}^2
/ \eta)^{-2} {\cal R_B} \;\;,\;\;
\frac{\nu_{c,r}(t_{\times})}{\nu_{c,f}(t_{\times})}={\cal
R_B}^{-3} \;\;,\;\;
\frac{F_{\nu,m,r}(t_{\times})}{F_{\nu,m,f}(t_{\times})}=(\gamma_{\times}^2
/ \eta) {\cal R_B}  \label{eq:evoforrev}
\end{equation}
where
\begin{equation}
\gamma_{\times}=\min(\eta, \eta_c)~~~,{\rm and}~~~ {\cal
R_B}\equiv B_r/B_f=
\left(\epsilon_{B,r}/\epsilon_{B,f}\right)^{1/2}.
\label{eq:pararev}
\end{equation}
Here $\cal R_{B}$ reflects a possible stronger B field in reverse
shock, as inferred from the analyses of the GRB 990123 and GRB 021200
data (Zhang et al. 2003). In our calculations we set ${\cal R_B}=1$ as
the standard case, and take ${\cal R_B}=5$ as an alternative option.
 As an example, when the observer time is larger than
the crossing time, $t \ge t_{\times}$=max($T,t_{dec}$), i.e. the
fast cooling case, the observed cooling frequency, typical
frequency and peak flux of the reverse shock are
\begin{eqnarray}
\nu_{c,r}& =&(t_{\times}/ t)^{3/2} {\cal R_B}^{-3}
\nu_{c,f}(t_\times)
=2.5\times10^{12}\epsilon_{B,r}^{-3/2}E_{52}^{-1/2}n^{-1} t_5^{-3/2}t_{\times,5}\times(1+z)^{-1/2} {\cal R_B}^{-3}\ Hz,\nonumber \\
\nu_{m,r}&=&\gamma_{\times}^{-4}\eta^{2} (t_{\times}/ t)^{3/2}
{\cal R_B} \nu_{m,f}(t_\times) =4.6
\times10^{14}\gamma_{\times}^{-4}\eta^{2} \epsilon_{B,r}^{1/2}
\epsilon_e^2
E_{52}^{1/2}t_5^{-3/2} \times (1+z)^{1/2} {\cal R_B} \ Hz,\\
F_{\nu,m,r}&=&\gamma_{\times}^2\eta^{-1}(t_{\times}/t) {\cal R_B}
F_{\nu,m,f}(t_\times) =1.1 \times 10^5
\epsilon_{B,r}^{1/2}E_{52}n^{1/2}D_{28}(z)^{-2}\times (1+z)
\gamma_{\times}^2\eta^{-1} t_{\times,5} t_5^{-1}{\cal R_B} \ \mu
Jy \nonumber. \label{eq:revafevo}
\end{eqnarray}

In the appendix we give further details of the expressions for the
flux evolution of the forward and reverse shocks in the thin and thick
shell as well as in the fast or slow cooling cases.


\subsection{GRB density environment }
The typical environments considered for GRBs are either the
(approximately) constant number density case $n_0 \sim$ constant
(i.e. independent of the distance $r$ from the center for the
burst), or a power law dependence as might be expected in the
stellar wind  from the progenitor, e.g. $n \propto r^{-2}$
(M\'{e}sz\'{a}ros et al. 1998; Dai \& Lu 1998; Chevalier \& Li
1999; Whalen, Abel, \& Norman, 2003). In our calculation, for
simplicity we consider only the first case of $n\sim$ constant,
which appears to satisfy most of the observed cases that have been
analyzed (Panaitescu \& Kumar 2001,2002; Frail et al. 2001). While
this density is different for different bursts, we can assume a
typical average value $n_0$ for $n$ at redshift $z=1$. One has to
consider then how this typical density might evolve with redshift.
We concentrate on two very different types of dependencies,
motivated by different physics. (1) Based on hierarchical models
of galaxy formation (Kauffmann, White, \& Guiderdoni 1993; Mo,
Mao, \& White 1998), the mass and size of galactic disks is
expected to evolve with redshift (Barkana \& Loeb 2000). For a
fixed host galaxy mass, this yields $n(z)=n_0 (1+z)^4$ (Ciardi \&
Loeb 2000). (2) Recent numerical simulations of primordial star
formation indicate that the particle number density around the
first stars at very high redshift could be in the range $1
\lesssim n_0 \lesssim 10^{-2}$ cm$^{-3}$ (Whalen et al. 2003),
approximately independent of redshift because of strong radiation
pressure from the central massive star, which dominates and
smooths any variations in the original galactic number density
around the stars. The size scale of this region of dominance is
about several parsecs, which is $\gtrsim$ the length scale of
typical afterglows. Here we assume that, for this case (2), this
stellar dominance applies to all GRBs originating from massive
stars, so the number density in the relevant region around the GRB
is the same constant at all redshifts, i.e.  $n(z)=n$. Thus, the
two density cases considered are
\begin{equation}
n(z)=n_0\times \left\{ \begin{array}{l@{\quad \quad}l}
 (1+z)^0 & \hbox{constant density model} \\
 (1+z)^4 & \hbox{density evolution model}
 \end{array} \right.
 \label{eq:densitypro}
\end{equation}
Here $n_0$ is normalized by $n_0=1$ cm$^{-3}$ at $z=1$,
noting that uncertainties in the primordial star calculations
could make this as low as $10^{-2}$ cm$^{-3}$.
This number density $n$ refers to the local ISM density
in the immediate neighborhood of the burst.

\subsection{Intergalactic and galactic absorption}

As it propagates through the intergalactic medium (IGM), the
afterglow radiation from a burst occurring at some redshift $z$ is
subject to several absorption processes. The most important are
Lyman-$\alpha$ absorption, photoionization of neutral hydrogen,
and photoionization of He II. At very high redshifts, before the
intergalactic medium becomes re-ionized, which may be taken to
occur between the limits $z_i \gtrsim 6.3$ (Fan et al. 2001;
Miralda-Escud\'e 2003; Onken \& Miralda-Escud\'e 2003) and $z_i
\sim 17 \pm 5$ (Spergel et al. 2003), most of the mass as well as
most of the volume of the IGM is in the form of neutral gas. At
redshifts below this, after re-ionization by the first stars or
galaxies, an increasing fraction of the IGM volume becomes
ionized, interspersed with clouds of neutral gas associated with
the halos of protogalaxies, which continue to absorb radiation.
The exact distribution of clouds as a function of redshift is not
well known, but estimates of the effective number are obtained by
counting the numbers of absorption line systems in quasar spectra.
These are used for calculating the effective absorption optical
depth at redshifts below the reionization redshift. Below the
reionization redshift, the photoionization opacity by HI is given
by Madau, Haardt, \& Rees (1999), based on the observed absorber
distribution in the spectra of high-redshift quasars. The
Lyman-$\alpha$ absorption optical depth can be obtained in a
similar way. Above the reionization redshift, both the
photoionization and Lyman-$\alpha$ opacities are obtained by means
of an integration through the neutral gas between the reionization
redshift and the redshift at which the GRB is located (Barkana \&
Loeb 2001).

At high redshifts, intergalactic He II becomes important at
rest-frame energies $\gtrsim 54.4$ eV, where the effects of
hydrogen photoionization are still important (Perna \& Loeb 1998).
However, the combined effect of the cross sections and the
abundances, as well as the hardness of the ionizing spectra
combine together to make He II the dominant opacity at observed
photon energies $h\nu\gtrsim 54.4 \hbox{eV}/(1+z)$ for sources
located at $z\gtrsim 3$ (Miralda-Escud\'e 2001). Bluewards of this
energy, as the cross section drops as $\nu^{-3}$, He II
photoionization is the last process to become optically thin, and
is therefore  the dominant IGM constituent which determines the
re-emergence of the source spectrum at frequencies above the blue
end of the Gunn-Peterson trough. Adopting current values of the
cosmological parameters, this occurs (M\'esz\'aros \& Rees 2003)
at soft X-ray energies of $h\nu_t\sim 0.2$ keV or $\nu_t\sim
5\times 10^{16}$ Hz.

Absorption by our own galaxy also becomes important in the UV and
soft X-ray band. The combined cross section including galactic
metals $\sigma_{ph}$ is given by Morrison \& McCammon (1983). The
optical depth is given by $\tau=\sigma_{ph}N_{H,Galaxy}$, where
$N_{H,Galaxy}$ is the equivalent column density along the line of
sight, which varies depending on the galactic latitude. Here we
set the column density to be $2 \times 10^{20}$ cm$^{-2}$, typical
of moderately high latitudes, which becomes optically thin at
energies $\gtrsim 0.2$ keV, comparable to the effects discussed
above for the intergalactic He II.

Thus, one expects that between the Lyman-$\alpha$ frequency corresponding to
the source frame and approximately $5\times 10^{16}$ Hz (below which the
galactic extinction for the above column density becomes large), the flux
observed from a high redshift GRB will be totally suppressed. Outside this
range, the observed flux is much less affected by the intergalactic and
galactic absorption.


\section{Initial Conditions and Numerical Results}

In our calculations, the nominal GRB parameter values adopted are
an isotropic-equivalent energy $E_{52}=10$, shock parameters
$\epsilon_e=0.1$, $\epsilon_{B,r}=0.025$, the ratio of magnetic
field strength in reverse and forward shocks ${\cal R_B}
=B_{r}/B_{f}=$ 1 or 5 ($\epsilon_{B,f}=0.025$ or
$\epsilon_{B,f}=0.001$, respectively), and an initial Lorentz
factor $\eta=120$. The GRB duration is assumed to be $T_s=10$ s in
the source frame. The deceleration time $t_{dec,s}=[(3E/4\pi
\eta^2 n m_p c^2)^{1/3}/2 \eta^2 c]$ in the source frame is
determined mainly by GRB intrinsic parameters, except for the
external ISM density $n$, which can depend on redshift in one of
the scenarios considered. Substituting the parameters for
$t_{dec,s}$, we have $t_{dec,s}=119.6 n^{-1/3} \hbox{seconds}$.
Therefore, for the $n$=const scenario, the reverse shock is
exclusively in the thin shell case; for $n \propto (1+z)^4$
scenario the reverse shock will be in the thin shell case below
some redshift, and above that redshift it will be in the thick
shell case. We take a specific case where the reionization
redshift of the Universe is at $z_i=15$, compatible with the WMAP
value of Spergel et al (2003). As examples, we considered the
burst properties at various observer's times, e.g. 10 minutes, 2
hours and 1 day. The results are presented in Figs.
\ref{fig:rflux_revfor_const} and \ref{fig:rflux_combined},
discussed below.

\subsection{Light Curve}

\begin{figure}[ht]
\epsscale{0.6} \plotone{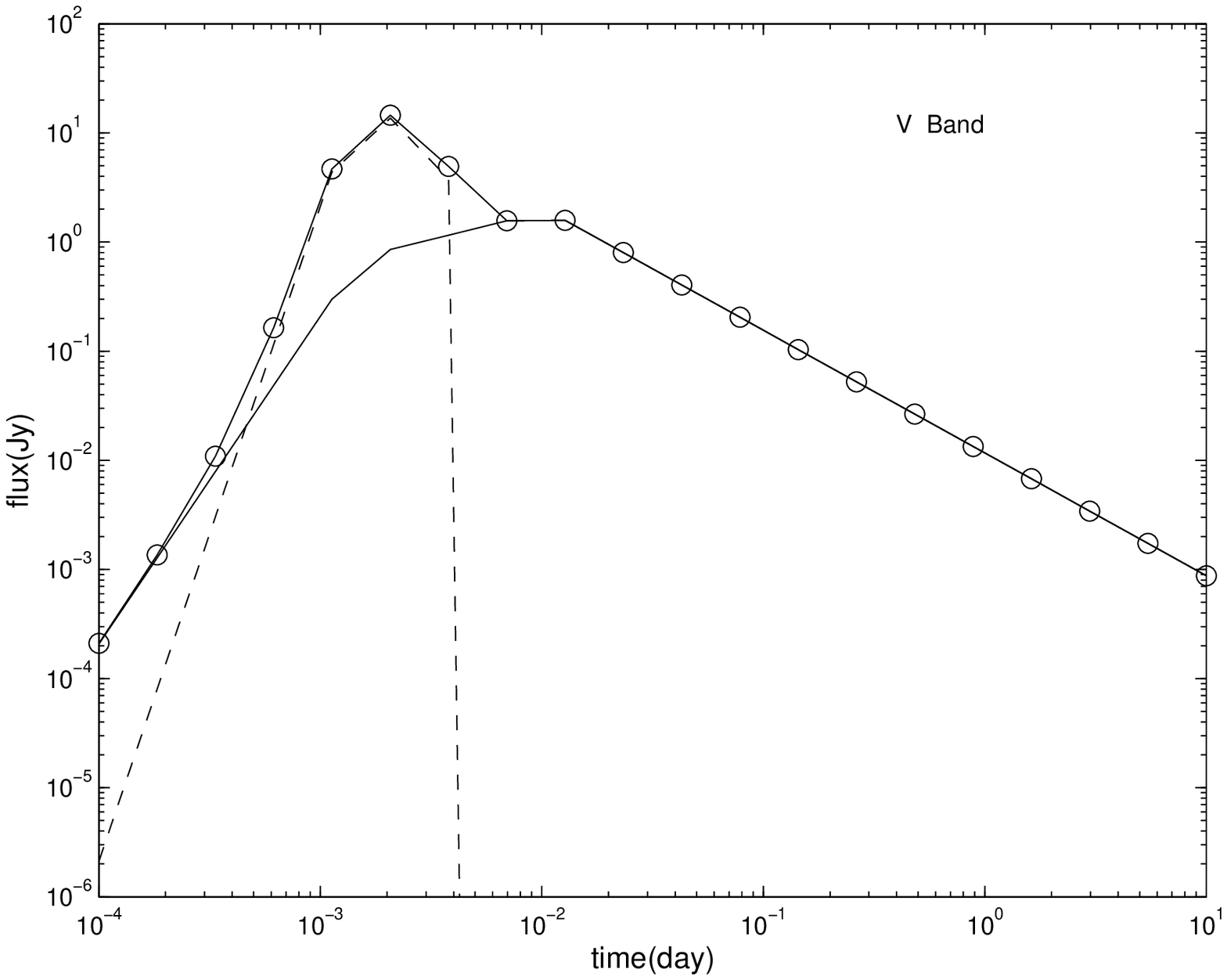} \plotone{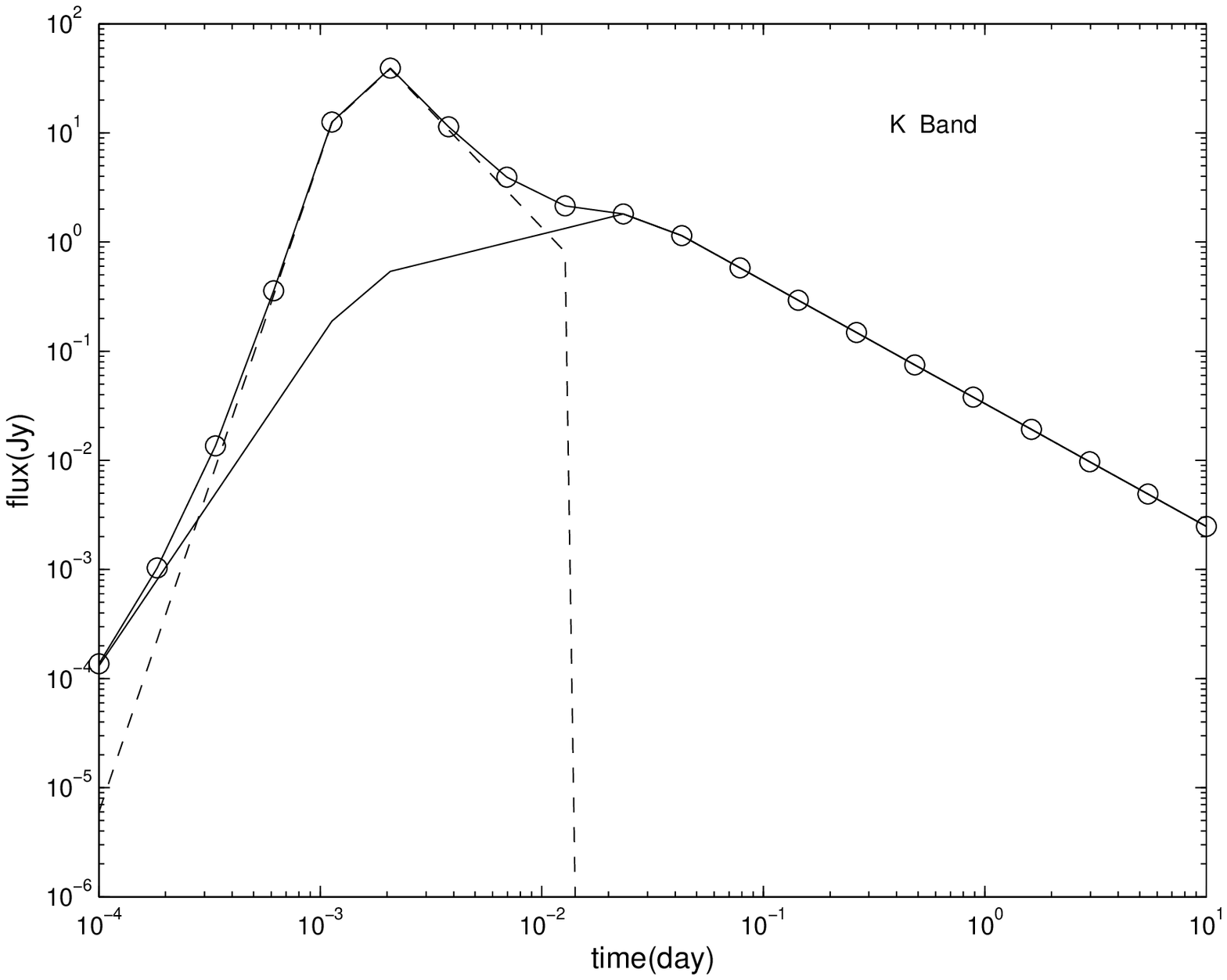}
\caption{Typical light curves, for a redshift $z=1$. Reverse shock
emission (dashed), forward shock emission (solid). total flux
(symbols). Parameters:$ \epsilon_{B,f}=0.001, {\cal R}_B=B_r/B_f=5,
\epsilon_e=0.1, E_{52}=10, p=2.5, \eta=120, n_0=1\ {\rm cm^{-3}}$.
a): V band ($\nu=5.45\times10^{14}$ Hz); b): K band
($\nu=1.36\times10^{14}$ Hz). } \label{fig:light_curve}
\end{figure}

To show the distinct effects of the reverse and the forward
shock on the flux behavior, we show the light curves for two
different observational bands, V and K, here taken at a nominal
redshift $z=1$. (Figure \ref{fig:light_curve}). As can be seen,
the light curve evolution can be divided into 3 stages: (1)
Forward shock dominant, before the reverse shock peaks. The
timescale for this stage is relatively short. (2) Reverse shock
dominant, after the reverse shock emission, which increases very
quickly, exceeds the forward shock emission. The flux peaks when
the reverse crosses the fireball shell.
(3) Forward shock dominant. In this stage, after the reverse shock
peak, the reverse shock emission decays very quickly and falls
below the forward shock, so the forward shock is again dominant.

\subsection{Infrared Flux Redshift Dependence}

\begin{figure}[ht]
\epsscale{0.5}
\plotone{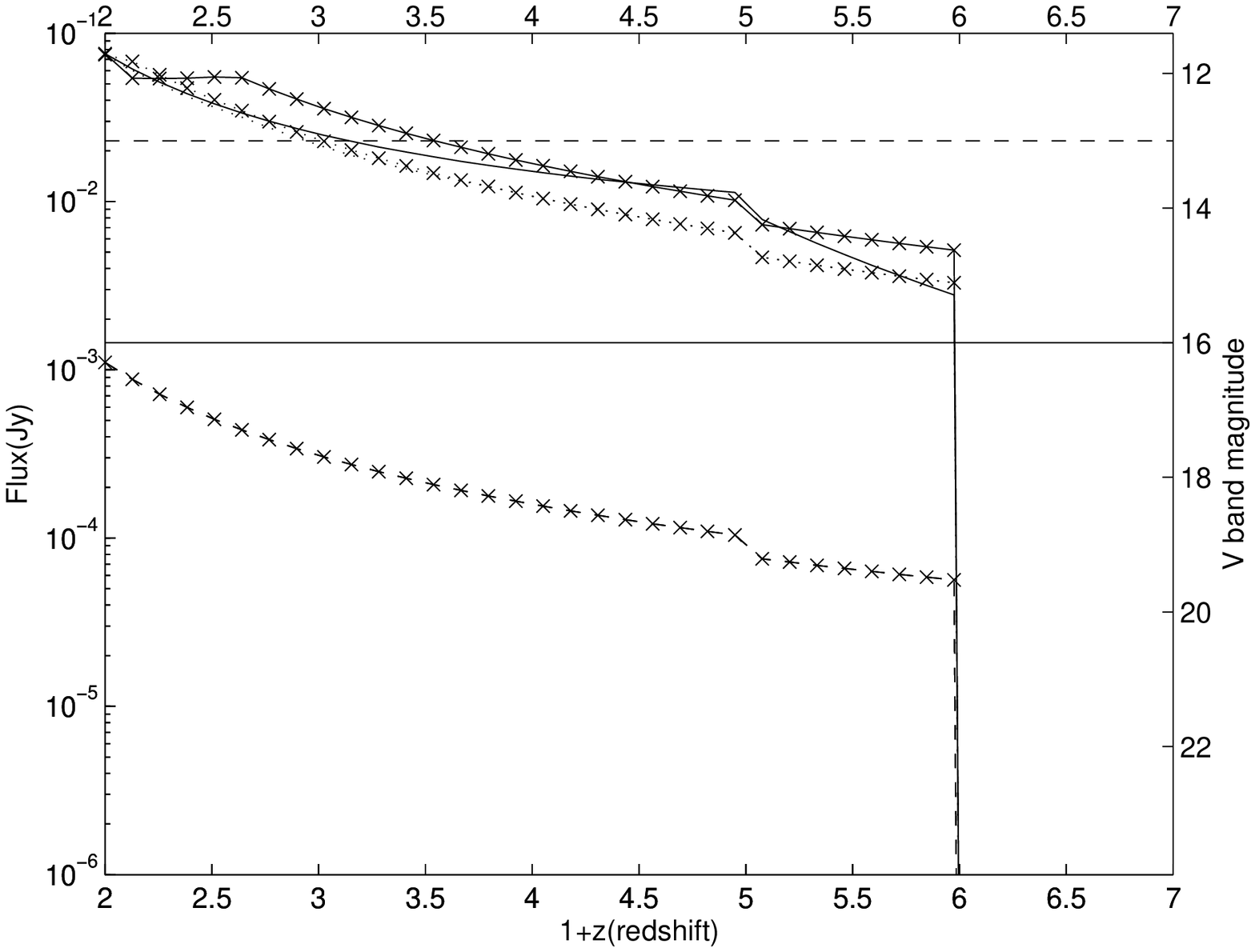}
\plotone{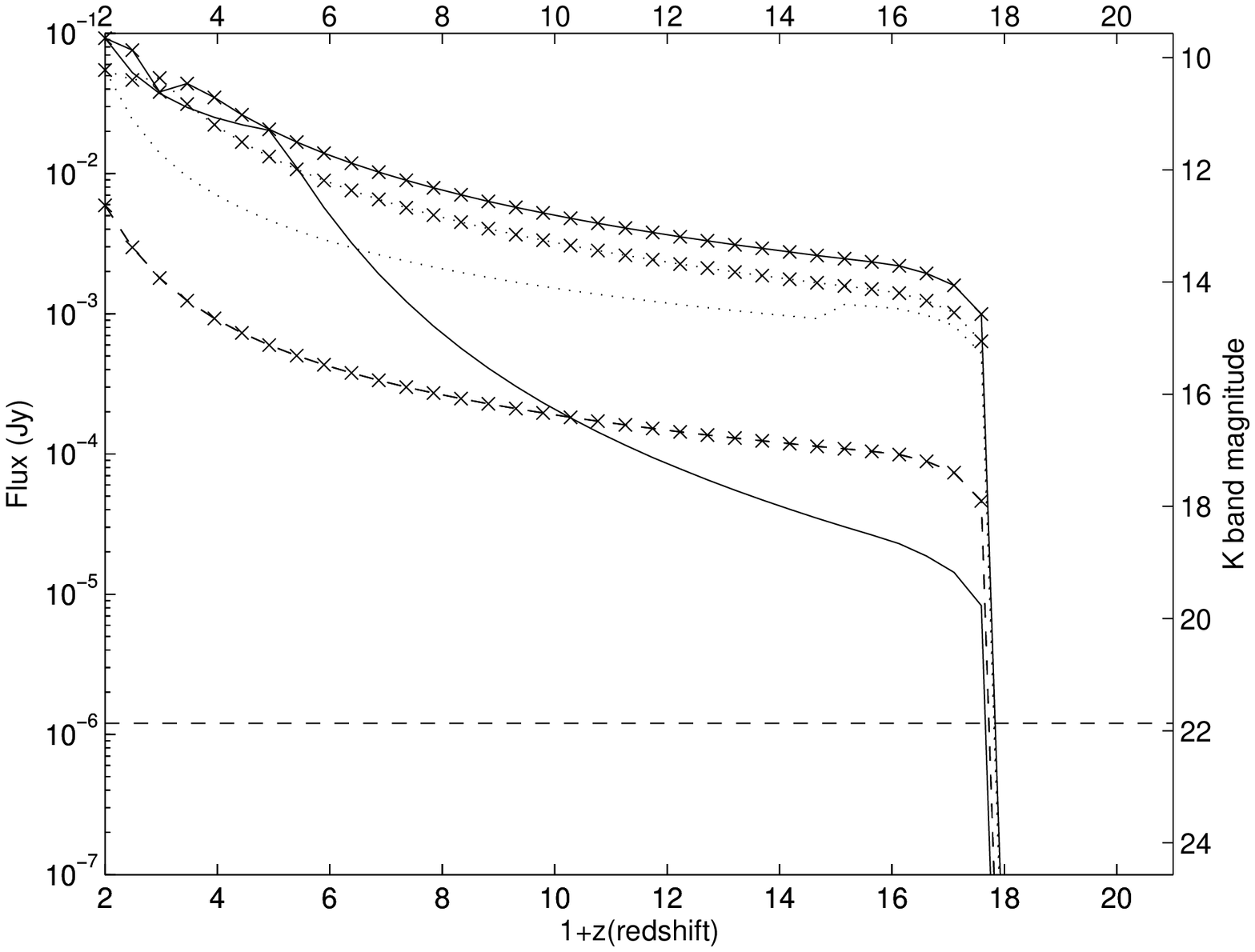}
\plotone{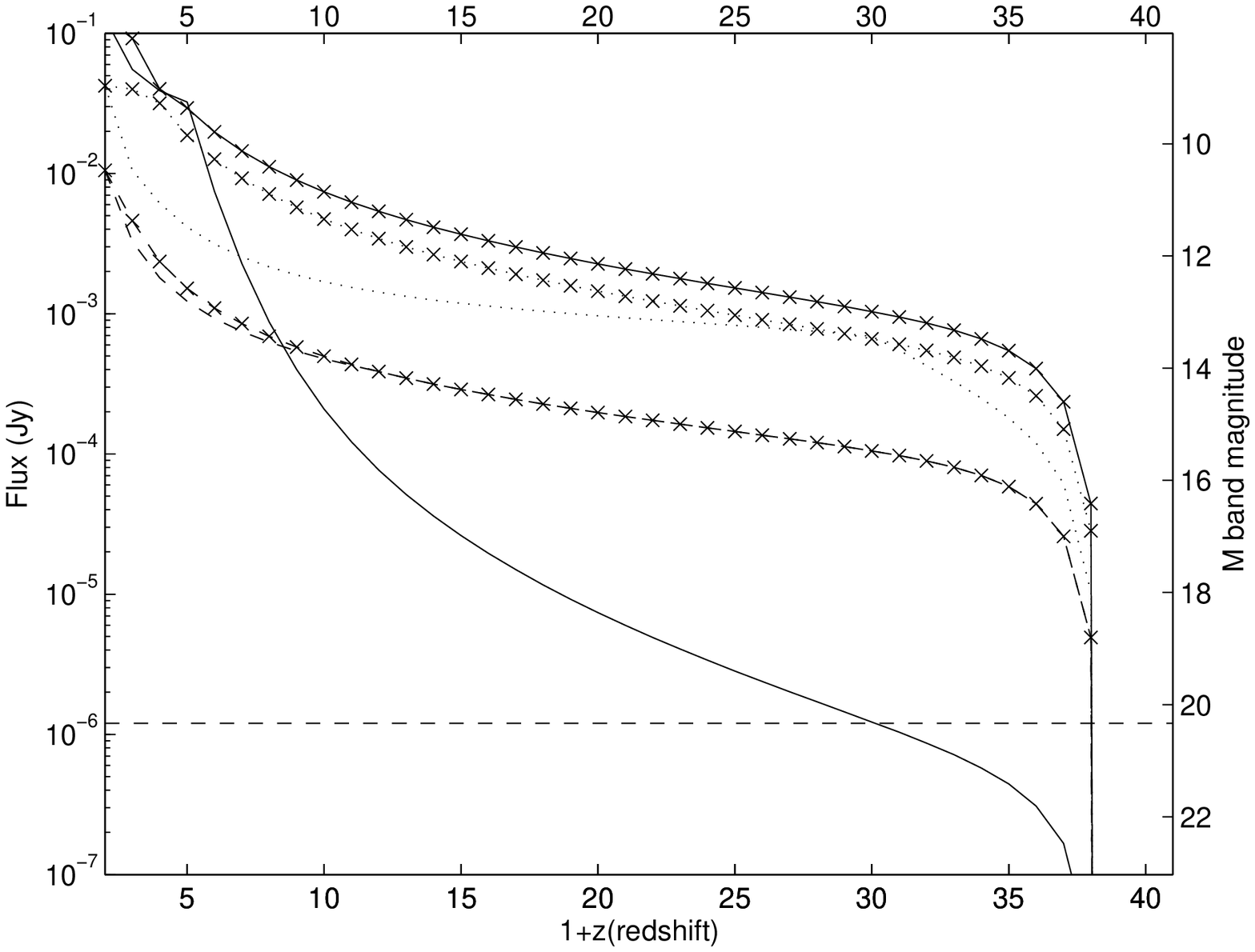}

\caption{ Combined forward and reverse shock observed flux as a
function of redshift for $\epsilon_{B,f}=0.025$ and ${\cal
R}_B=B_r/B_f=1$. Forward shock (symbols), reverse shock (without
symbols). Solid, dashed and dotted lines indicate emission at
different observer times t=10\ mins, t=2\ hour and t=1\ day
respectively. a): V-band ($\nu=5.45\times10^{14}$ Hz); b): K-band
($\nu=1.36\times10^{14}$ Hz); c): M-band ($\nu=6.3\times 10^{13}$
Hz). Straight lines: In V band sensitivities are for ROTSE at very
early and late times; in K and M bands sensitivities for JWST K \&
M bands are estimated for a resolution R=1000, S/N=10 and
integration time of 1 hour. Parameters: $n=1 \ {\rm cm^{-3}},
\epsilon_e=0.1, E_{52}=10, p=2.5, \eta=120.$}

\label{fig:rflux_revfor_const}
\end{figure}

\begin{figure}[ht]
\epsscale{0.5}
\plotone{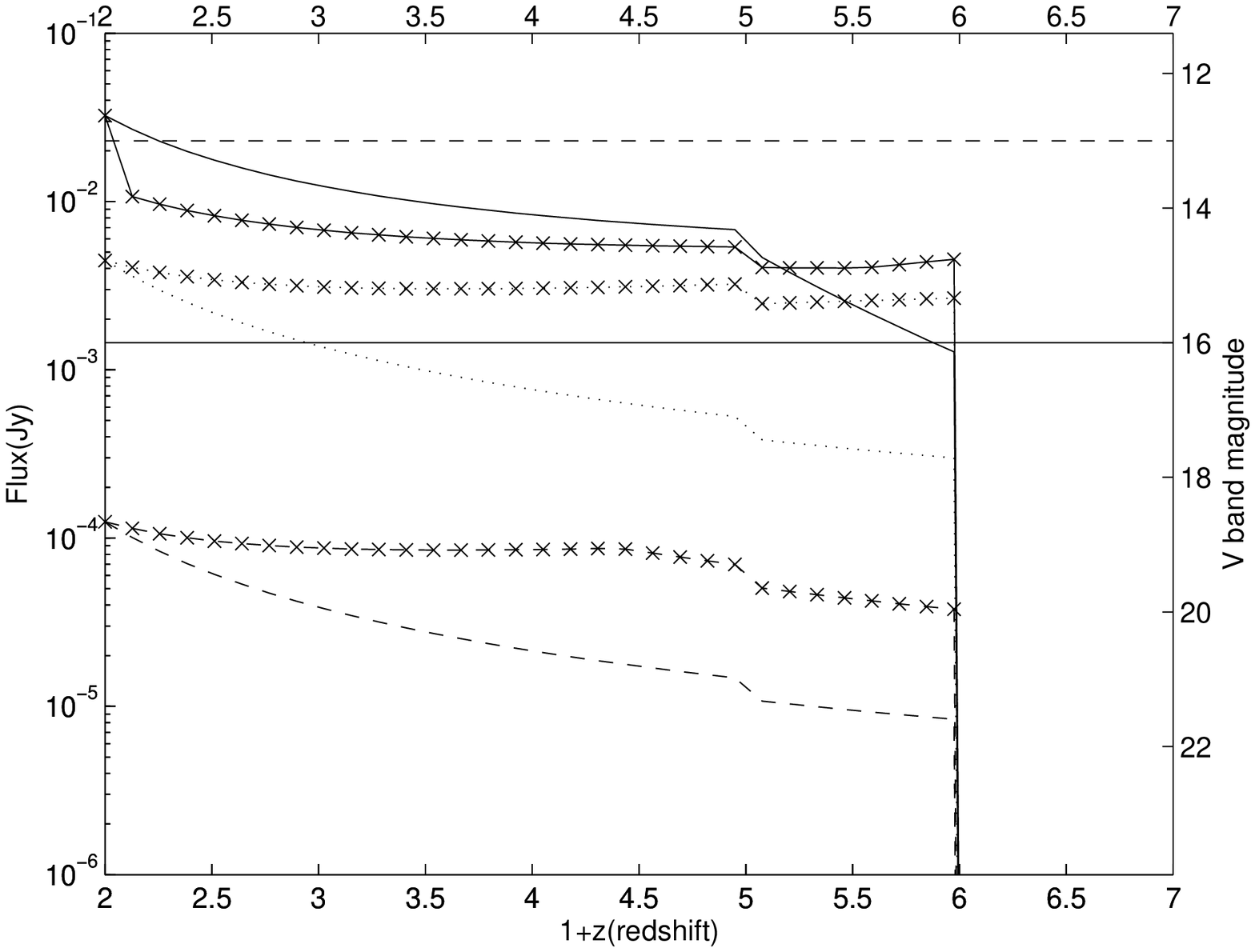}
\plotone{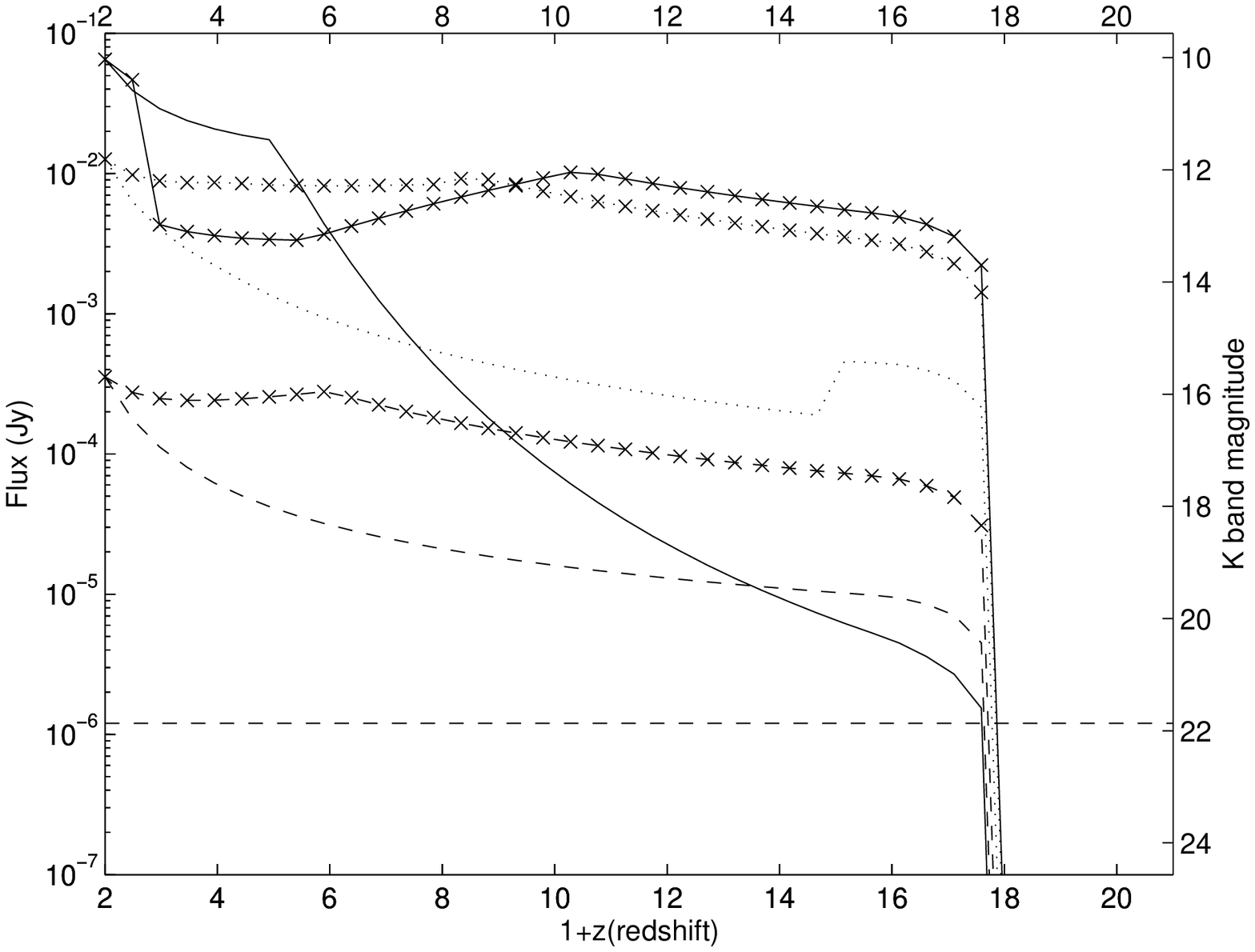}
\plotone{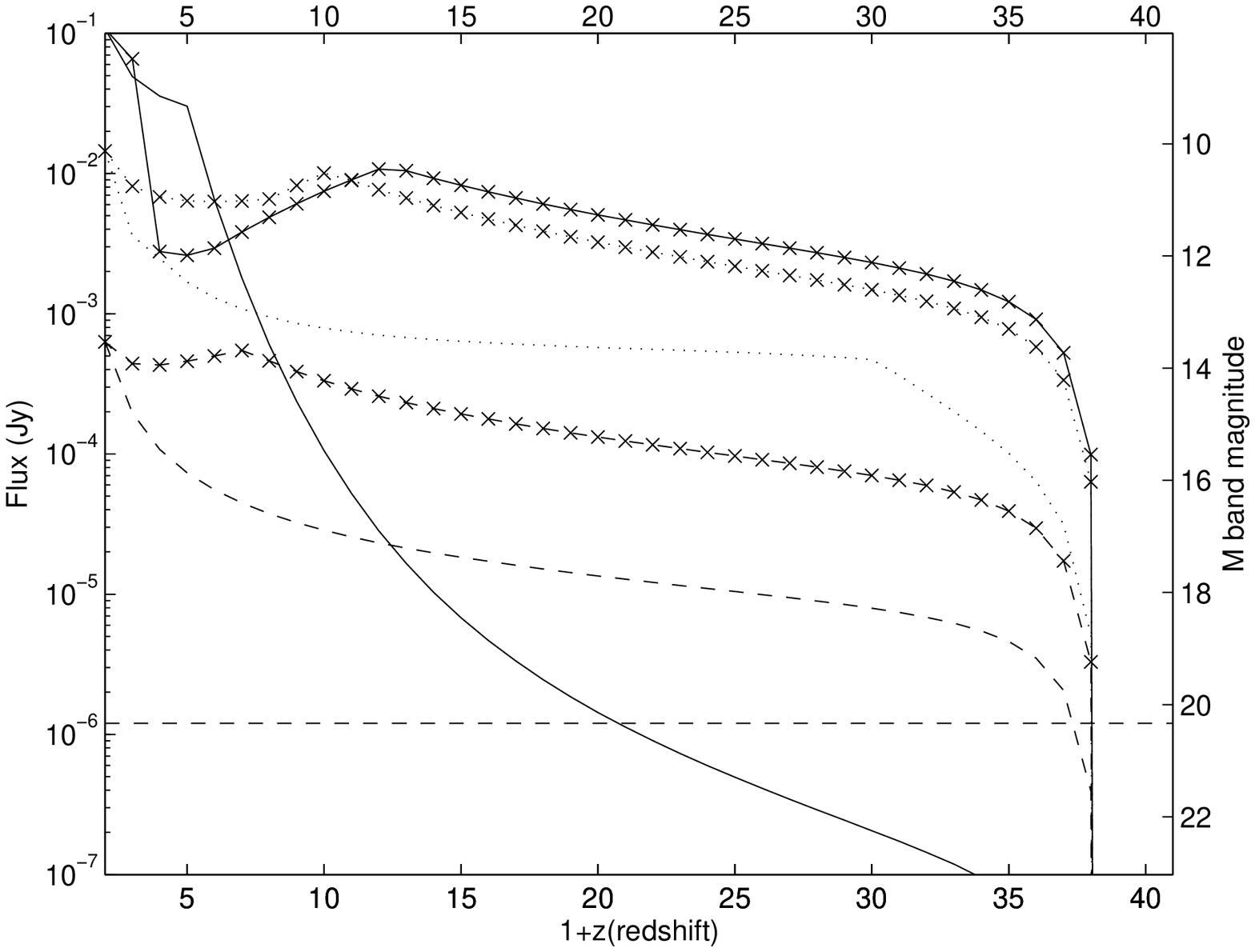}

\caption{ Combined forward and reverse shock observed flux as a
function of redshift for the two density profiles $n=n_0=$
constant ( without symbols) and $n=n_0 (1+z)^4$ (symbols) with
$n_0=1$ cm$^{-3}$. solid, dashed and dotted lines indicate
emission at different observer times t=10\ mins, t=2\ hour and
t=1\ day respectively. a): V-band ($\nu=5.45\times10^{14}$ Hz);
b): K-band ($\nu=1.36\times10^{14}$ Hz); c): M-band
($\nu=6.3\times 10^{13}$ Hz). The limiting ROTSE and JWST
sensitivities are the same as in Figure
\ref{fig:rflux_revfor_const}. Parameters: $ \epsilon_{B,f}=0.001,
{\cal R}_B=B_r/B_f=5, \epsilon_e=0.1, E_{52}=10, p=2.5, \eta=120.
$} \label{fig:rflux_combined}
\end{figure}

To test the self consistency of the code with the present optical
observation like ROSTE, we plot the light curves in V band
(Figures \ref{fig:rflux_revfor_const}(a) and
\ref{fig:rflux_combined}(a)). On these two figures the dashed an
solid lines correspond to the sensitivities of ROTSE at very early
and late times, respectively. The light curves show that we don't
expect to see many optical flashes, especially for high-$z$ GRBs.
This is consistent with the present observations.

From the flux evolution equations (see appendix), it is seen that
in the regime where the observing frequency is above the cooling
frequency, $\nu>\nu_{c,f}$,  the observed flux for the forward
shock component is independent of the ISM number density (see
(\ref{eq:scalforfast}) \& (\ref{eq:scalforslow})). We can define a
critical redshift $z_c$, such that for $z>z_c$ the GRB afterglows
for the forward shock component are in the density-independent
regime (see eq.(3)).
\begin{eqnarray}
(1+z_c)& =& 3.4 \times 10^{2} (\epsilon_{B,f}/0.01)^{-3} E_{52}^{-1} n^{-2} t_5^{-1}(\lambda/2.2 \mu \hbox{m})^2 \nonumber \\
& = &1.1 \times 10^{2} (\epsilon_{B,f}/0.01)^{-3} E_{52}^{-1}n^{-2}
t_5^{-1}(1 \hbox{eV}/h\nu)^{-2} \label{eq:zc}
\end{eqnarray}
From the above equation (\ref{eq:zc}), we see that the dependence
of $z_{c}$ on $\epsilon_{B,f}$ is very sensitive, $\propto
\epsilon_{B,f}^{-3}$. If we set $\epsilon_{B,f}=0.1$, we obtain the
equations given by Ciardi \& Loeb (2000). Taking $\epsilon_{B,f}$
smaller, the redshift $z_c$ can increase substantially. That is
one of the main reasons why our curve of flux vs. redshift differs
from that of Ciardi and Loeb (2000).

For the description of reverse shocks there are four relevant cases,
depending on whether one is in the thin or thick shell limit, and on
whether the times considered are before or after the shock crossing time.
However, the cooling frequency evolution can be approximated by
$\nu_{c,r} \propto t^{-3/2}$ for observer times
$t > t_{\times}=\max(t_{dec}, T)$.
Also, if the observed frequency is larger than the cooling frequency
$\nu \ge \nu_{c,r}$, the reverse shock emission disappears.
Hence, we can define another critical redshift $z_r$ at which the
reverse shock emission disappears,
\begin{equation}
(1+z_r) = 3.33 \times 10^{-6} (\epsilon_{B,r}/0.01)^{-3} E_{52}^{-1} n^{-2}
 (t_{\times}/10\hbox{s})^2 t_5^{-3}(\lambda/2.2 \mu \hbox{m})^2 {\cal
 R_B}^{-6}
\label{eq:zr}
\end{equation}
which is a lower limit for $n \propto (1+z)^4$ and is an upper limit for
$n=n_0=$constant, i.e.  for $z \ge z_r$ when $n \propto (1+z)^4$ or for
$z \le z_r$ when $n$=const, there is no reverse shock emission (see below).

The two critical redshifts can be connected by the relations
$\nu=\nu_{c,r}=(t_{\times}/t)^{3/2}{\cal R_B}^{-3}
\nu_{c,f}(z_{r})$ and $ \nu=\nu_{c,f}(z_{c})$. Cancelling out
$\nu$ and substituting the expression for $\nu_{c,f}$, we obtain
the relation
\begin{equation}
(1+z_r)=(t_{\times}/t)^3 {\cal R_B}^{-6}(1+z_c),
 \label{eq:eqzrzc}
\end{equation}
and we have the inequality
\begin{equation}
z_r<z_c
\label{eq:ineqzrzc}
\end{equation}
since ${\cal R_B}\geq 1$, and $t>t_\times$ by default, i.e. $z_r$ is
defined for $t>t_\times$.

Using the parameters above and an observer frequency
$\nu=1.36\times 10^{14}$ Hz ($6.3\times 10^{13}$ Hz) or
$\lambda=2.2\ \mu$m ($4.8\ \mu$m), corresponding to the K-band
(M-band) at observer times $t=10$ mins, 2 hour and 1 day, we have 
(${\cal R_B}=1$)
\begin{equation}
z_r= \cases{2.2,~ 0.1,~  0 ~ (3.0,~ 0.4, ~0)  & for $n \propto
(1+z)^4$ ; \cr 0,~   120.6,  ~\infty ~(0, ~24.5,~ \infty) & for $n=1
~ \hbox{cm}^3$   . \cr} \label{eq:zrcases}
\end{equation}

From equation (\ref{eq:revafevo}) we have, for the $n \propto
(1+z)^4$ case,
\begin{equation}
\nu_{c,r,evolv}=2.5
\times10^{12}\epsilon_{B,r}^{-3/2}E_{52}^{-1/2}n_{z=0,evolv}^{-1}
t_5^{-3/2}t_{dec,s,5}\times (1+z)^{-7/2}{\cal R_B}^{-3} \hbox{Hz}
\propto (1+z)^{-29/6},
 \label{eq:crevolv}
\end{equation}
and for the $n=const$ case
\begin{equation}
\nu_{c,r,const}=2.5 \times10^{12}\epsilon_{B,r}^{-3/2}E_{52}^{-1/2}
n_{z=0,const}^{-1}t_5^{-3/2}t_{dec,s,5}\times (1+z)^{1/2}{\cal
R_B}^{-3} \hbox{Hz} \propto (1+z)^{1/2}.
 \label{eq:crconst}
\end{equation}

From equations
(\ref{eq:zr}),(\ref{eq:crevolv}),(\ref{eq:crconst}), we find some
interesting differences between the two density profile cases. We
discussed that for $\nu > \nu_{c,r}$ there is no emission from the
reverse shock. However, the above behavior of $\nu_{c,r,evolv}
\propto (1+z)^{-29/6} $ for $n \propto (1+z)^4$ and
$\nu_{c,r,const} \propto (1+z)^{1/2} $ for the $n=n_0=$ constant
case has some other consequences.  For the same burst parameters,
if there is no reverse shock emission at some observer time $t$
for $z=0$, in the $n \propto (1+z)^4$ case this implies that there
will be no reverse shock emission at this same observer time at
any redshift. For the $n=$ constant case, however, the chances are
that the reverse shock emission will be observable above some
redshift, because $\nu_{c,r,const}$ increases with redshift. This
is caused by the effect of time dilation increasing as the
redshift increases, which means that the same observer-frame time
corresponds to earlier and earlier source-frame times. In the $n
\propto (1+z)^4$ case, $\nu_{c,r,evolv} \propto (1+z)^{-29/6} $,
and we can expect that $\nu_{c,r}$ will decay with $z$ quickly
below $\nu$ even if $\nu_{c,r}$ is much larger than $\nu$ at low
redshift. So, in this case, we can only observe the reverse shock
emission at relatively low redshifts. On the other hand, in the
$n=$ constant case we can observe the reverse shock emission at
all redshifts if there is emission at low redshifts. We can see
this from the fluxes in figure \ref{fig:rflux_revfor_const}. If we
substitute the values for the relevant parameters, we get $z\le
z_r=2.1$ and $z\le z_r=0$ at $t=10$ mins and $t=1$ hour
respectively for the $n\propto (1+z)^4$ case, whereas for the
$n=const$ case, $z\ge z_r=0$ and $z\ge z_r=1975$ for those two
corresponding times, and the emission from the reverse shock is
observable in K band. This property provides one way of
distinguishing these two different density profile regimes, based
on the redshift distribution of the occurrence or absence of a
reverse shock component.

From equations (\ref{eq:eqzrzc},\ref{eq:ineqzrzc}) we see that if
$t>t_{\times}$  for the $n \propto (1+z)^4$ case, the reverse
shock emission is absent already at redshifts lower than those
beyond which the GRB emission would be in the density-independent
regime. On the contrary, for the $n$=const case, reverse shock
emission exists when the GRB emission is in the
density-independent regime. Using this characteristic, we can
again constrain the density profile around the GRBs.

Looking at Figure \ref{fig:rflux_revfor_const} which is the
standard case of ${\cal R_B}=1$ in our calculation, several
features can be noted: (1) At early times, e.g. $t=10 \ {\rm
mins}$ and $t=2$ hour, we can differentiate the constant density
profile from the evolving density profile in both K and M
bands. However, at late times it becomes difficult to do so in both
bands, although the total flux in M band are somewhat
different for both density profiles at relatively low redshifts. (2)
For early observer times, e.g. $t=10\ {\rm mins}$, the amplitude of
the total flux in the evolving density $n\propto (1+z)^4$ case at low
redshifts shows some pronounced and complicated changes with
redshift, as opposed to a more monotonous behavior in the constant
density case. The changes in the former are caused by the
transitions from one regime to another by the forward shock. Because
in the evolving density example reverse shock emission
disappearance above some very low redshift, over the redshift
considered  here forward shock emission component is dominant.
On the other hand, in the constant density case, over the entire
redshift range considered reverse shock emission is dominant. (3) The
break in the light curve for the $n=$ constant case at $t=10$ minutes
is caused by the transition from $t >t_{\rm dec,s}$ to $t<t_{\rm
dec,s}$. (4) There is a sharp decline in the emitted flux in light
curves at redshift $z\sim~17$ for K band and at $z\sim36$, which are
caused by the lyman-$\alpha$ and photoionization absorption of
neutral hydrogen in IGM.

We also considered the ${\cal R_B}=5$ case (Figure
\ref{fig:rflux_combined}) which indicates that magnetic field in
the reverse shock is much stronger than that in the forward shock.
Other parameters in this case are same as those in Figure
\ref{fig:rflux_revfor_const}. The most distinct feature for this
case from the standard case is that there is one jump around
redshift $z\sim2$ in the evolving density case. This jump is caused
by the disappearance of reverse shock above some redshift. Based
on the equations
(\ref{eq:thinbefluxfast}--\ref{eq:bothafterslow}), the flux ratio
between reverse and forward shock increases with an increasing
${\cal R_B}$. So when the reverse shock emission disappears,
we can expect a sudden jump in the light curve as a function of
redshift.

\begin{figure}[ht]
\epsscale{0.5}
\plotone{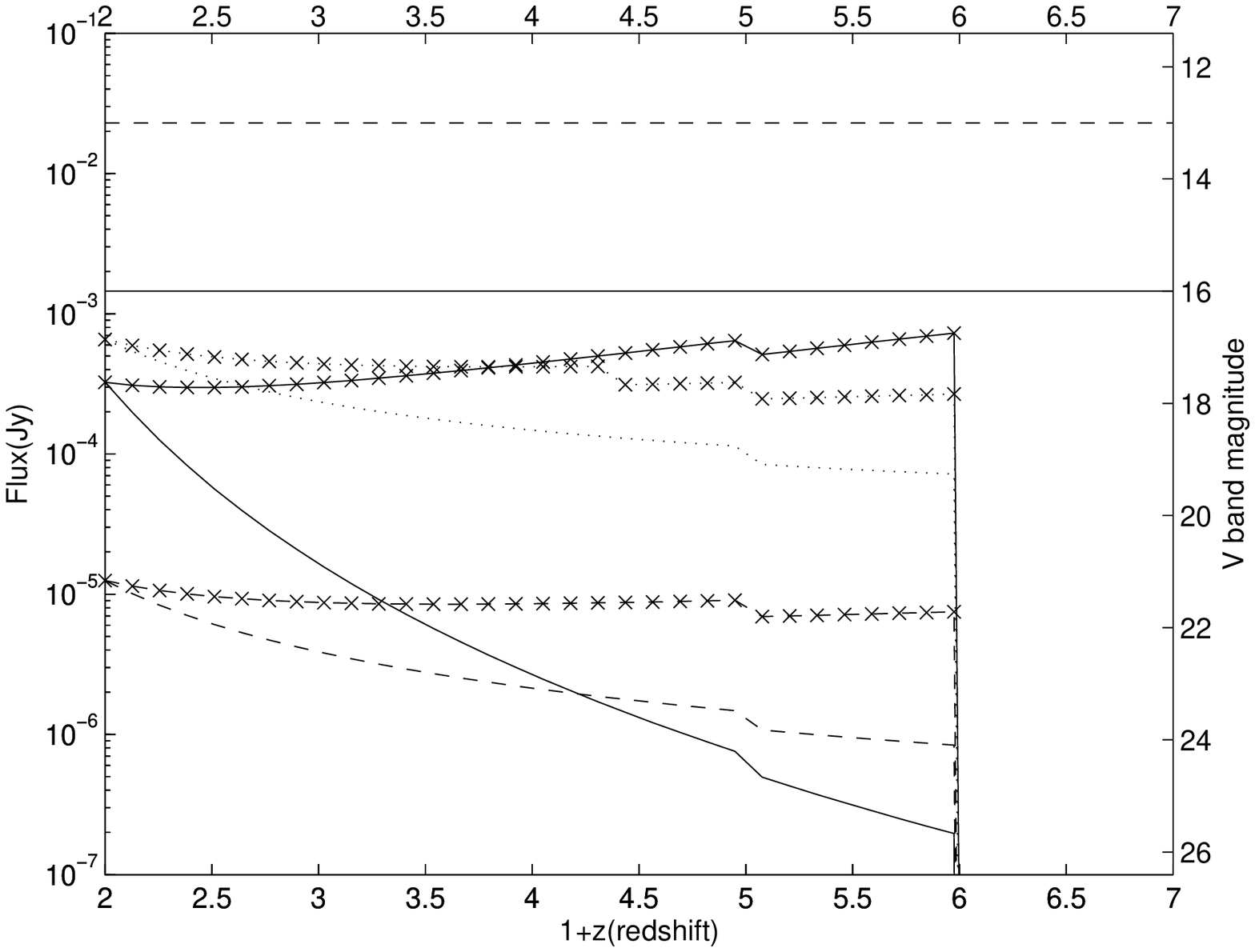}
\plotone{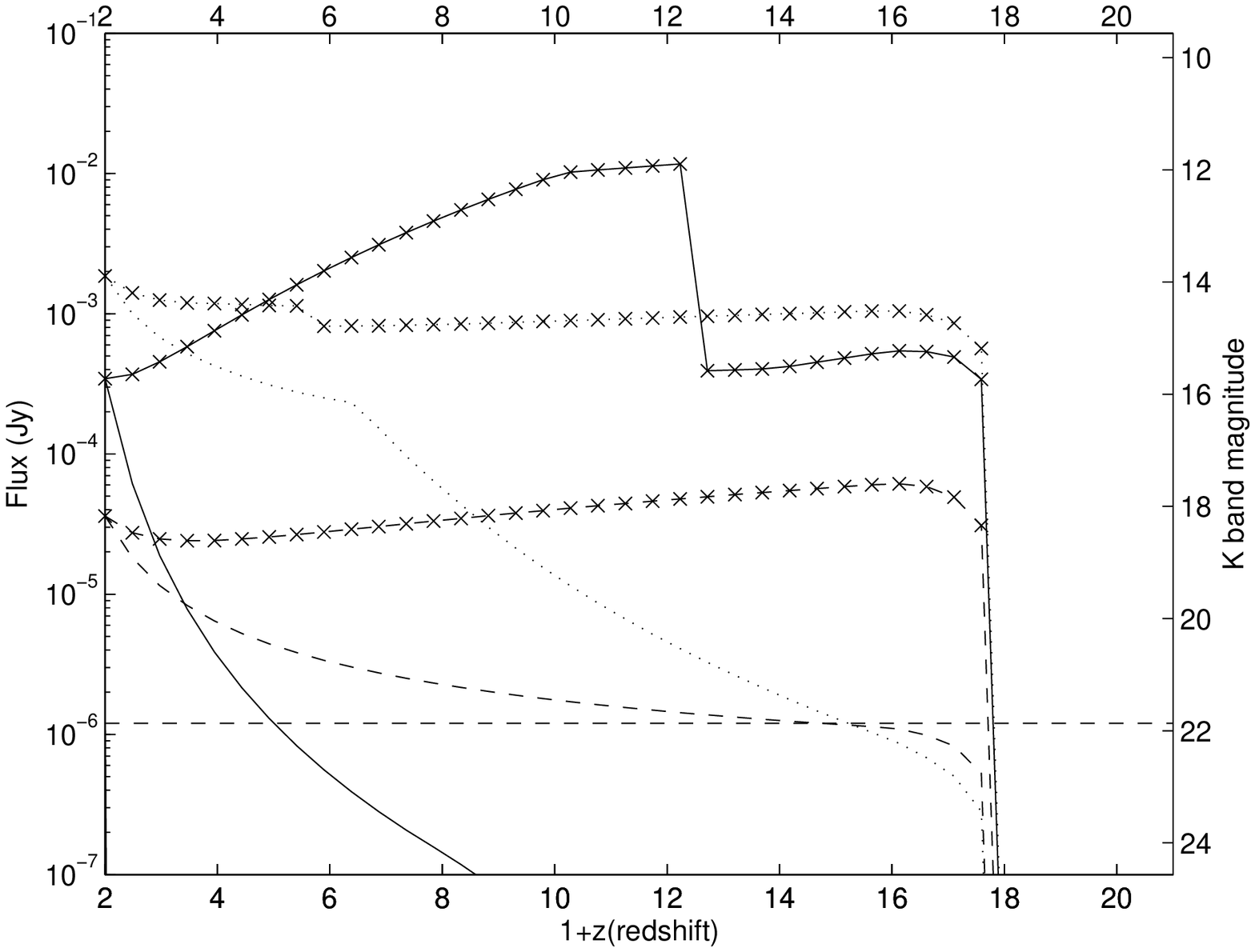}
\plotone{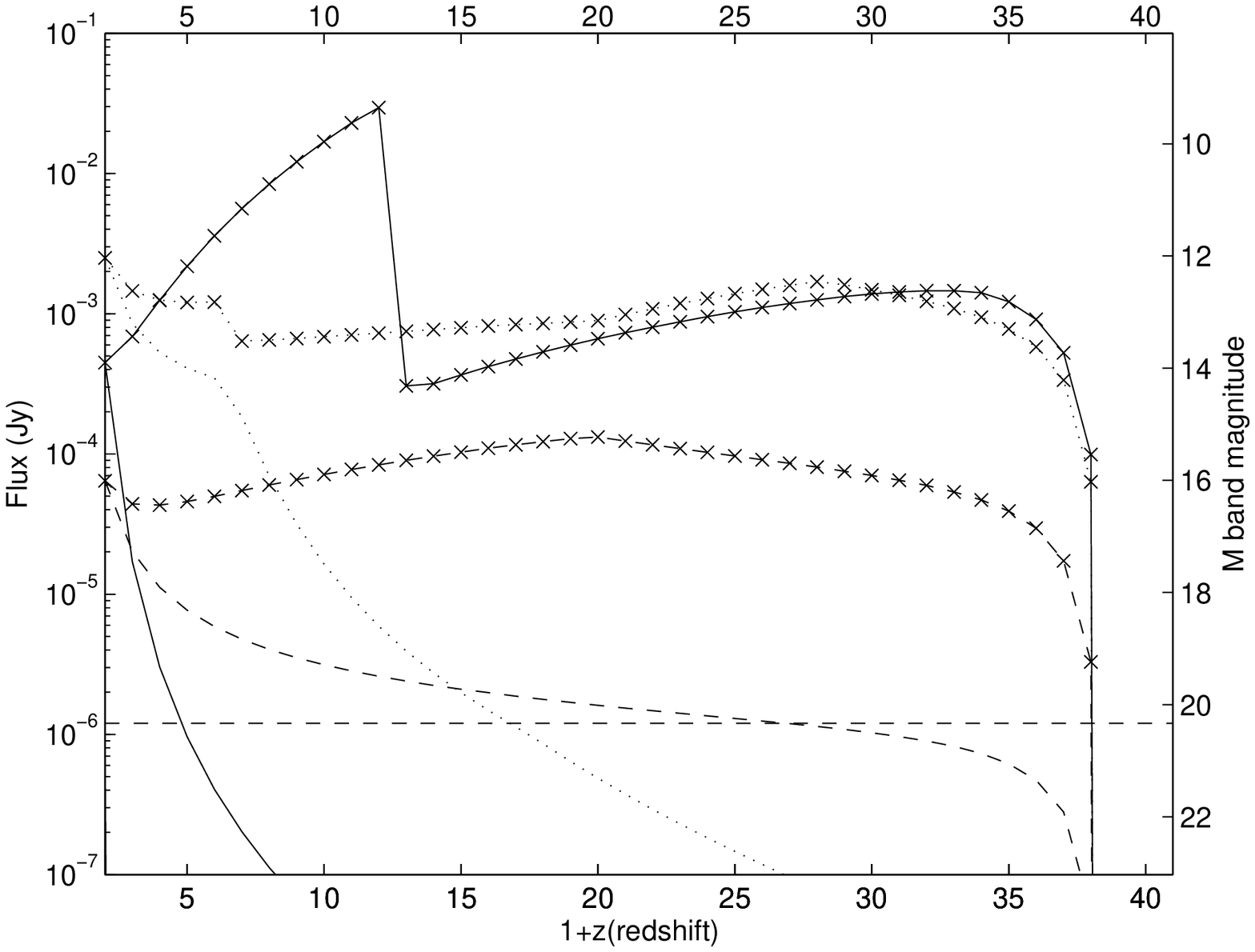}

 \caption{ Combined forward and reverse shock
observed flux as a function of redshift for the two density
profiles $n=n_0=$ constant (without symbols) and $n=n_0 (1+z)^4$
(symbols) with $n_0=0.01$ cm$^{-3}$. Solid, dashed and dotted
lines indicate emission at different observer times t=10\ mins,
t=2\ hour and t=1\ day respectively. a): V-band
($\nu=5.45\times10^{14}$ Hz); b): K-band ($\nu=1.36\times10^{14}$
Hz); c): M-band ($\nu=6.3\times 10^{13}$ Hz). The limiting ROTSE
and JWST sensitivities are the same as in Figure
\ref{fig:rflux_revfor_const}. Parameters: $ \epsilon_{B,f}=0.001,
{\cal R}_B=B_r/B_f=5, \epsilon_e=0.1, E_{52}=10, p=2.5,
\eta=120.$} \label{fig:rflux_low_den}
\end{figure}

We further tested a different normalization, i.e. $n_0=0.01\ {\rm
cm^{-3}}$, still for ${\cal R_B}=5$. The same total forward plus
reverse shock flux for the two density profiles for this case are
shown in Figure \ref{fig:rflux_low_den}. An obvious feature is that
the reverse shock is much more prominent for the lower
normalization density ($n_0= 0.01 \ {\rm cm^{-3}}$) case than for the
higher normalization density ($n_0=1 \ {\rm cm^{-3}}$) case, in the
density evolution model. From equations (\ref{eq:scalforfast},
\ref{eq:scalforslow},
\ref{eq:scthinaffast} \& \ref{eq:scthinbeslow}), for slow cooling,
the ratio of reverse shock to forward shock emission is
proportional to $n^{(1/6-1/2)}=n^{-1/3}$ or
$n^{(1/3-1/2)}=n^{-1/6}$. Therefore, the reverse shock emission
becomes more
prominent for a decreasing number density. However, we also note
that the reverse shock emission is smaller than the forward shock
flux in some cases. This is because the reverse shock is in the
$\nu_{m,r} < \nu< \nu_{c,r}$ regime before the crossing time. So
the ratio of the reverse shock to forward shock flux is
proportional to $n^{(3p+1)/4}$. When the number density is smaller
than 1 $\rm cm^{-3}$, we can expect the reverse shock flux to be
smaller. The same is also expected for the constant density
case at an early observer time. So in this case these two density
models can be easily differentiated from each other.

\subsection{X-ray Flux Redshift Dependence}

\begin{figure}[ht]
\epsscale{0.45}
\plotone{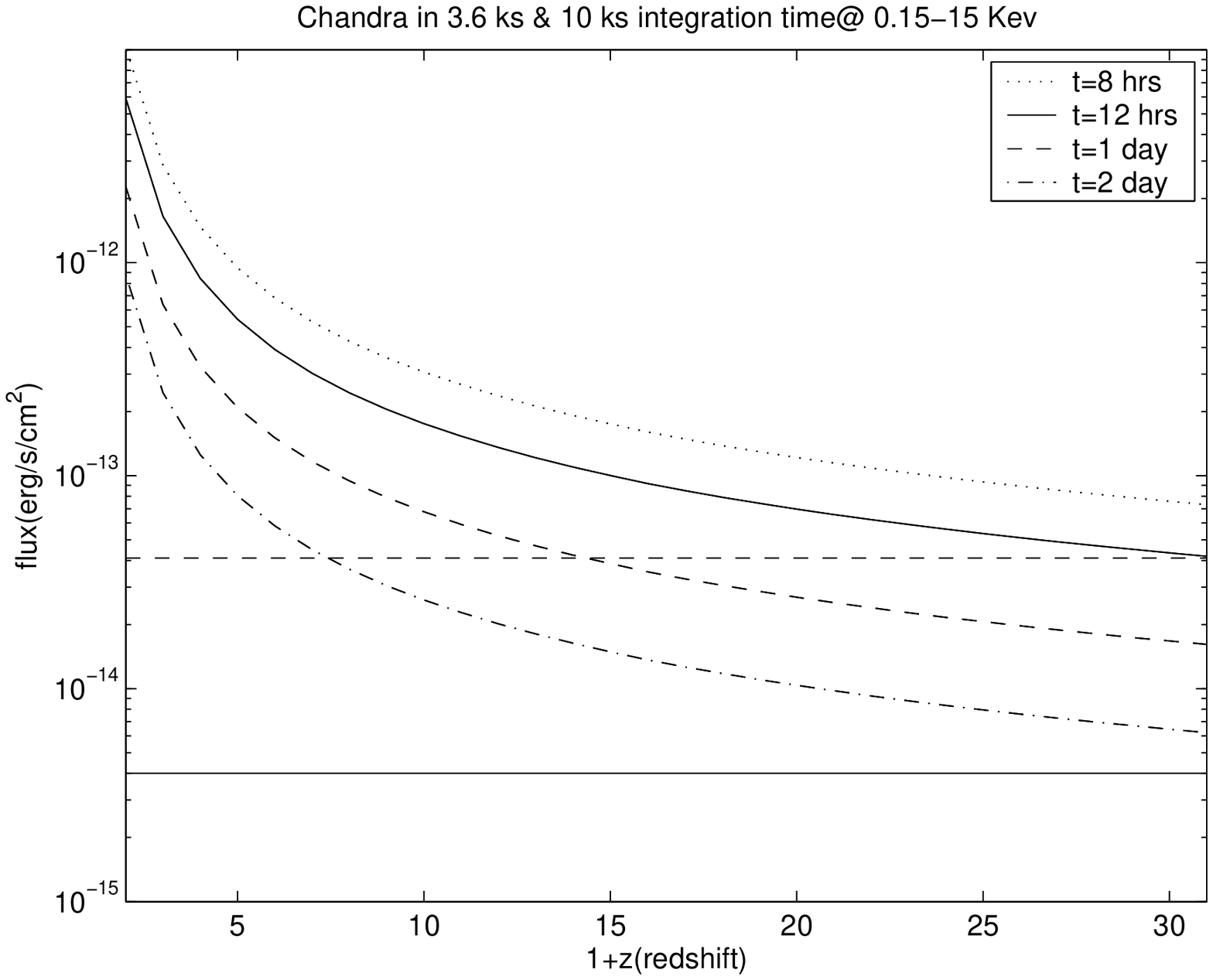}
\plotone{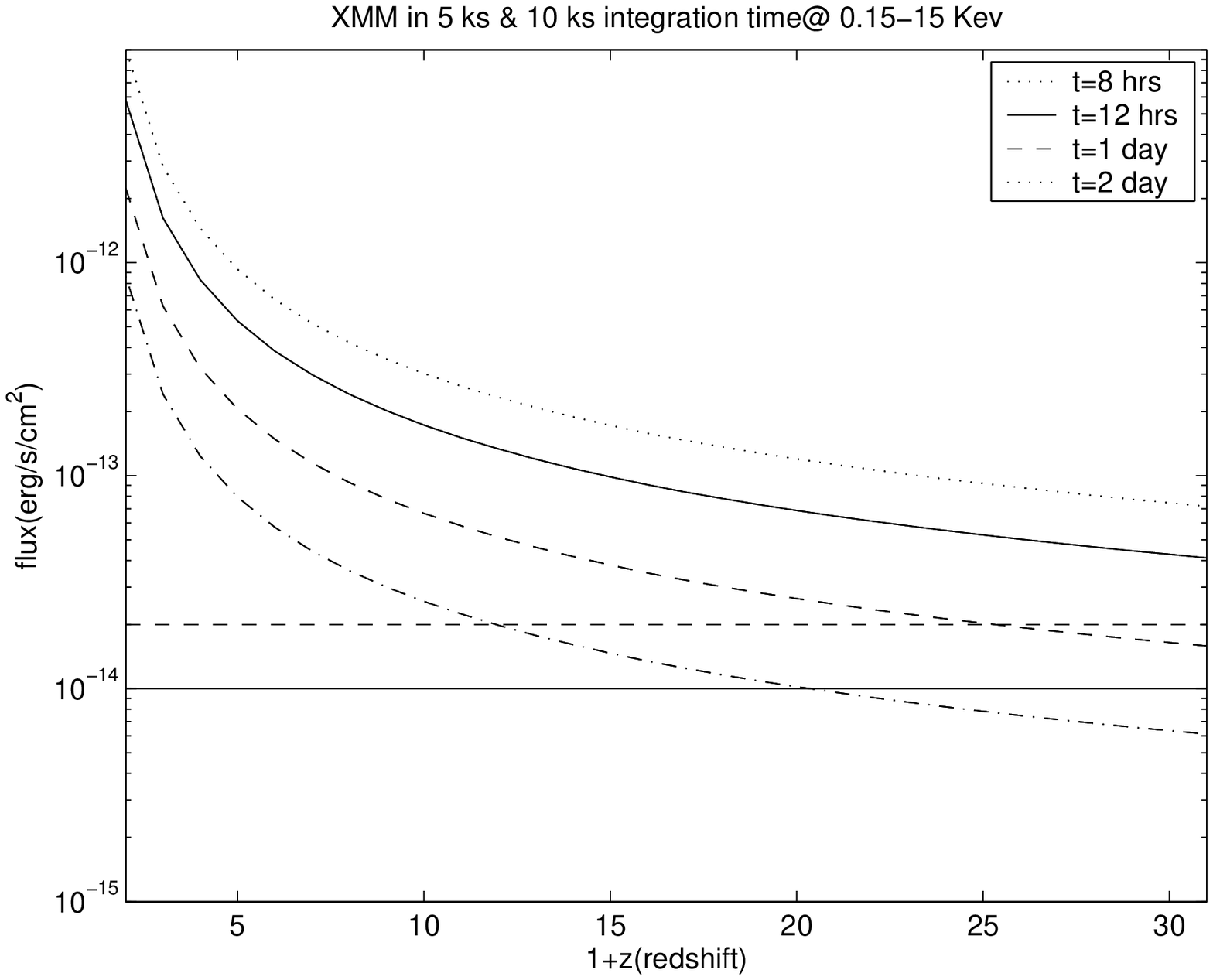}
\plotone{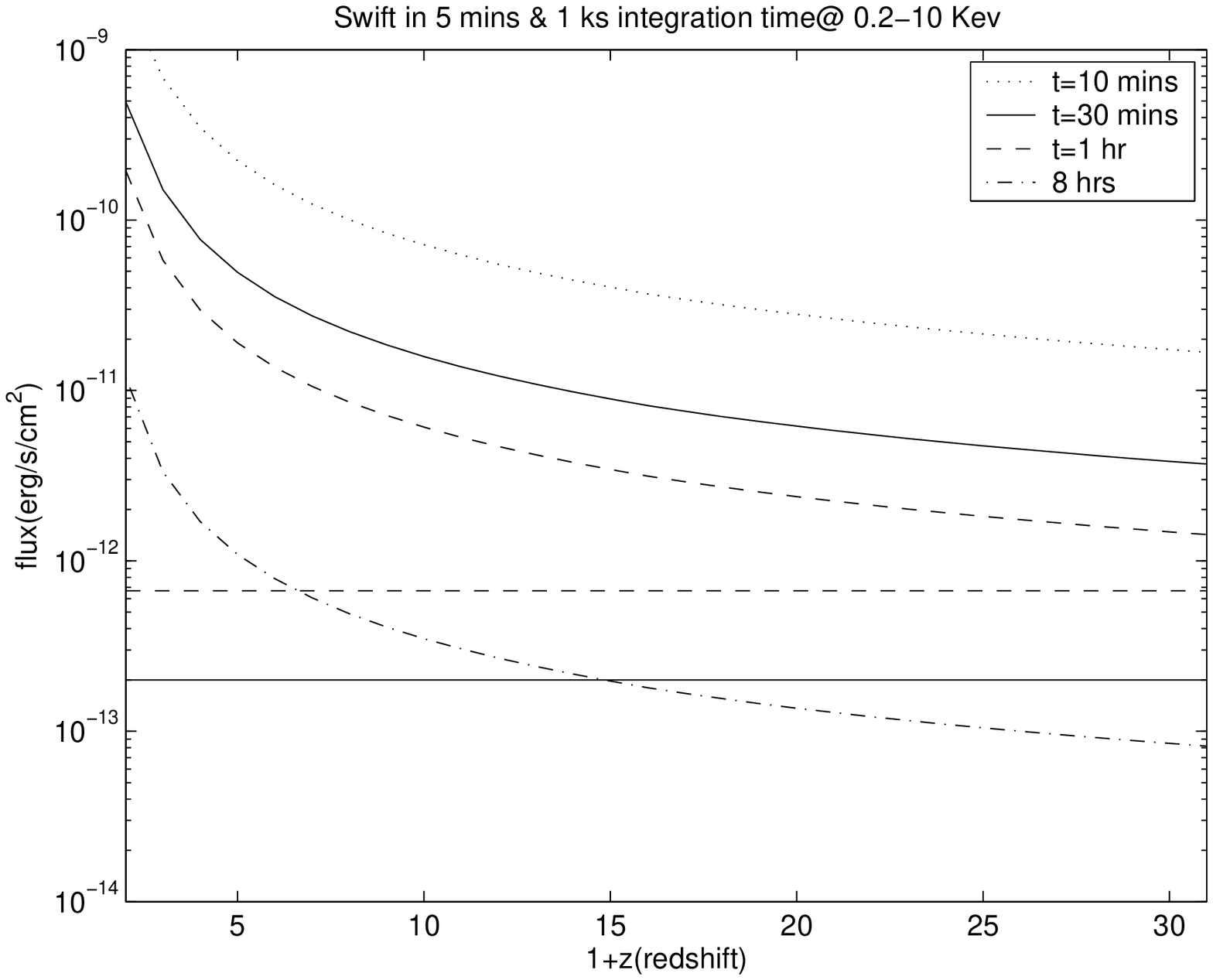}

\caption{Observed X-ray fluxes for GRB afterglows at different
redshifts, integrated over the observing energy ranges of 0.4-6
keV for Chandra, 0.15-15 Kev for XMM and 0.2-10 keV for Swift,
respectively. The emission is in the density-independent regime,
above $\nu_{c,f}$. a): for Chandra, the fluxes for the observer
times $t_{obs}=$ 8 hour, 12 hours, 1 day and 2 days as compared to
its sensitivities shown as horizontal lines for integration times
of 3.6 ks (dashed) and 10 ks (solid). b): for XMM, same observer
times as Chandra's. The sensitivity horizontal lines are for
integration times of 5 ks (dashed) and 10 ks (solid). c): for
Swift XRT, the fluxes are for observer time $t_{obs}=$ 10 mins, 20
mins, 1 hour. The sensitivity horizontal lines are for integration
times 300 s (dashed) and 1 ks (solid). Parameters: $
\epsilon_{B,f}=0.001, {\cal R}_B=B_r/B_f=5, \epsilon_e=0.1,
E_{52}=10, p=2.5, \eta=120.$} \label{fig:xrflux}
\end{figure}

The X-ray band flux evolution and its redshift dependence is
simpler than in the O/IR bands because the reverse shock emission
is generally negligible, and we need only consider the forward
shock emission. One obvious characteristic of Figure
\ref{fig:xrflux} is that the flux from the two different density
profiles are the same at all the redshifts for a given time,
because the emission in both cases is in the density-independent
regime, being above the cooling frequency $\nu_{c,f}$. Based on
equation (\ref{eq:zc}), for an arbitrary choice of low X-ray
energy of 0.1 keV and using the other parameters above, we obtain
a critical redshift $z_c = 0.83$ (n=const) or $z_c=0.98 \
(n\propto (1+z)^4)$ where the GRBs change from the
density-dependent to the density-independent regime at an observer
time $t=10$ mins. In addition, we know that $(1+z_c) \propto
t^{-1}\nu^{-2}$ based on eqn (\ref{eq:zc}). As the observer time
or observer's frequency is increased, the critical redshift
decreases, if the other parameters remain unchanged. Therefore, in
the range of redshifts concerned, GRBs in these two different
density profiles are always in the density-independent regime and
the corresponding fluxes will always be same. On Figure
\ref{fig:xrflux} (a) and (b), the X-ray flux is calculated for
observer times $t=8\ {\rm hours}, \ 12\ {\rm hours}, \ 1\ {\rm
day}$ and $2\ {\rm days}$ for Chandra and XMM, respectively. The
flux is integrated over the $0.4-6$ keV range of the Chandra ACIS
instrument, and over 0.15-15 kev for XMM. Although it takes about
1 day or so for Chandra to slew onto the source, we see that
Chandra is still able to detect GRBs with the typical parameters
considered here up to $z \approx 30$ at 1 day with a 10 ks
integration. XMM has proved itself capable of $\lesssim \ 8 \ {\rm
hours}$ slewing onto GRBs and has a similar sensitivity to
Chandra, hence it might be able to detect higher redshift GRBs
than Chandra does. In Figure \ref{fig:xrflux}(c), we have
integrated the X-ray flux over the Swift XRT frequency range of
$0.2-10$ keV. Although Swift XRF has a relatively lower
sensitivity than Chandra and XMM, this is compensated by its quick
slewing time, less than 1 minute. In Figure \ref{fig:xrflux}(c) we
see that Swift XRT can easily detect typical GRBs up to redshifts
$\sim 30$, if they are observed within 1 hour after the trigger.
Therefore, if GRBs exist at very high redshift, we can expect
these detectors to be able to measure them in x-ray band.


\section{Summary and Conclusions}

In this paper, we have calculated the spectral time evolution and
the flux in the near-IR K- and M-bands as well as in the X-ray
band from GRBs at very high redshifts and different times. In
previous work, Ciardi \& Loeb (2000) calculated IR fluxes as a
function of redshift and time for the standard forward shock model
of afterglows.  Lamb \& Reichart (2000) discussed the optical/IR
as well as gamma-ray fluxes of some observed bursts at an observer
time of 1 day when placed at different redshifts. Here we have
introduced several new elements in our analysis, motivated by
recent developments in the observations as well as in the modeling
of bursts. The most important of these is that we consider, in
addition to the forward shock, also the reverse shock, which has
now been inferred in three GRBs from prompt follow-ups. The quick
response capability of a number of ground and space observing
facilities coming on-line in the near future means that one is far
likelier to observe the early stages of the GRB and of its
afterglow evolution. Thus there are excellent prospects for
observing the reverse shock thought to be responsible for the
prompt optical flashes, which is prominent only at early times in
the burst evolution. The observation of reverse shocks will, in
addition, provide significant independent information on early GRB
evolution, such as the initial Lorenz factor, the strength of
magnetic fields, etc. (e.g. Zhang et al.
2003). Another difference with previous flux calculations is that
we have used significantly updated model parameters, based on new
data acquired in the past two years. Thus, for instance, we make
use of the emerging consensus view that $\epsilon_{B,f}$ is
usually smaller than $\epsilon_{e,f}$. As is seen from the
calculations presented here, the magnetic field equipartition
value has a significant effect on the flux.

We have calculated the flux from high redshift GRBs taking
typical parameters, which gives a sense for how far GRBs can be
detected using current or forthcoming instruments. In reality,
these parameters would vary over a wide range, which would also
affect the detectability. Zhang et al. (2003) have
reexamined the three cases of GRB 990123, GRB 021004 and GRB
021211 and found evidence for an enhancement of the magnetic field
in the reverse shock over that in the forward shock, by as much as
a factor of ${\cal R}_B=B_r/B_f \approx 15$ (GRB 990123), i.e.
$\epsilon_B$ in reverse shock is much larger that in the forward
shock. If so, based on eqns.
(\ref{eq:thinbefluxfast}-\ref{eq:bothafterslow}), we can see that
if $\cal R_B$ increases, the observed (optical/IR) flux for the
reverse shock component increases significantly as illustrated by
comparing Figures (\ref{fig:rflux_revfor_const}) and
(\ref{fig:rflux_combined}). Separately, it has recently been
recognized that there are burst to burst variations in the total
beaming-corrected energy, and thus, the forward shock peak
afterglow fluxes may also be significant (Bloom, Frail, \& Kulkarni
2003). The so-called `fGRBs' exhibit rapid, jet-break like decays
at early times, before the $\lesssim 1$ day point at which they
would usually be expected. Thus we can expect that the reverse
shocks in this kind of GRBs are less bright.

Most of the radiation in the optical and ultraviolet bands from
high redshift extragalactic sources, including GRBs, is absorbed
by the intergalactic medium and the diffuse gas in our own galaxy.
The X-ray and infrared bands are therefore of major importance for
detecting and tracking high-redshift GRBs. Several major
ground-based telescopes as well as smaller robotic facilities have
or will have infrared sensitivity in the K, L and M-bands. Next
generation spacecraft such as Swift have X-ray and optical/UV
detectors, while the James Webb Space Telescope (JWST) frequency
range extends out to 27 $\mu$m, being most sensitive in the 1-5
$\mu$m J,H,K,L and M bands. In the X-ray band, the Chandra and XMM
sensitivities in 0.2-10 keV are substantially higher than that of
Swift XRT, but their slewing time ($\lesssim 1$ day) limitations
make Swift XRT a unique instrument for X-ray follow-up during the
first day after a GRB trigger, when the burst is brighter. In
spite of this, all three spacecraft should be able to detect very
distant GRBs, if they exist, e.g. at $z \gtrsim 30$.

For the nominal GRBs considered here, the luminosities are comparable
to those of the currently detected ones. According to theoretical
modeling, the fractional number of GRBs expected
at $z\gtrsim 5$ is $\gtrsim 50\%$ of which $\sim 15\%$ may be
detectable in flux-limited surveys such as Swift's (Bromm \& Loeb
2002). It was reported that HETE-2 sees 13 out of 14 GRB optical
afterglows by Sep. 2003, which means that the high-$z$ GRB fraction
is small (Ricker 2003). However, recently another high-z GRB
candidate (GRB 031026) was detected and proposed (e.g. Atteia et al.
2003), which increases the high-$z$ fraction to be close to 15$\%$.
We note also that at $z\gtrsim 6-10$ the first generation
of (pop III) stars are likely to lead to black holes with masses
$10-30~M_\odot$, and hence to GRBs whose luminosities could be
factors 10-30 times higher (e.g. M\'esz\'aros \& Rees 2003) than
assumed here. In this case, the fraction at $z\gtrsim 5$
detectable in Swift's flux-limited survey could be $\gtrsim
20-30\%$ of the total, or $\gtrsim 20$/year.

In the K and M bands (2.2 and 4.8 $\mu$m) the JWST and other
telescopes should be able to detect afterglows out to $z \lesssim
16$ and 33 within observer times 1 day for integrations times
(with JWST) of 1 hour (at a resolution R=1000 and S/N=10; see
Figure \ref{fig:rflux_combined} and Figure
\ref{fig:rflux_low_den}). These bands are accessible also to
ground-based telescopes already on line, before the JWST launch.
The effect of reverse shocks, which are brighter in the O/IR at
early source times for some bursts, makes for a significantly
increased sensitivity at high redshifts at observer times
$\lesssim$ 1 day.

We have also considered the effect of two different types of the
near-burst environment, one assuming that the external density
evolves with redshift similarly to that in protogalactic disks,
and the other assuming approximately redshift-independent
conditions regulated by radiation pressure, based on primordial
star formation calculations. The predicted X-ray fluxes, being due
mainly to forward shocks above the cooling frequency, are
independent of the external density regime, as well as insensitive
to the existence of reverse shocks. However the IR fluxes are
sensitive to which of these density regimes prevails, and at early
times are very sensitive to the presence and strength of a reverse
shock component, in particular at early times and redshifts
$\lesssim 15$. Combining these two types of IR and X-ray flux
information will thus provide very important tools for detecting
GRBs (if present) out to very high redshifts, for studying their
local environments, and for investigating the effects of reverse
shocks as well as the prompt phases of the bursts and their
afterglows.

\acknowledgments We are grateful to the referee for useful comments,
to Shiho Kobayashi, George Chartas, David Burrows, 
Jian Ge and Zheng Zheng for useful discussions,
and to  NASA NAG5-9153, NAG5-9192, NAG5-13286  
grants and the Monell Foundation  for support.

\appendix

\section{Appendix: Flux Evolution}

\subsection{Forward Shock}

The radiation emitted by a source at a redshift z at frequency
$\nu_s$ over a time $\delta t$ will be observed at z=0 at a
frequency $\nu_0(z=0)=\nu_s/(1+z)$ over a time $\delta
t_{0}(z=0)=(1+z)\delta t_s$. The luminosity distance for a flat
universe $\Omega_{\Lambda}+\Omega_M=1$, $\Omega_M=0.27$ and Hubble
constant $(H/100)\ km/s/Mpc=0.7 h_{70}$ can be approximated, in
units of $10^{28}$ cm, as $D_{28}(z)\approx
4.49(1+z)[1-1.115(1+z)^{-1/2}]$ (Pen 1999). Substituting this
redshift dependence and eqn. (\ref{eq:fnum}) into equations
(1) and (2), we have for the fast cooling case
\begin{equation}
F_{\nu}=\left\{ \begin{array}{l@{\quad \quad}l}
C_1\epsilon_{B,f} E_{52}^{7/6}t_5^{1/6}n^{5/6}\nu^{1/3}(1+z)^{-5/6}x^{-2} & \nu<\nu_{c,f} \\
C_2\epsilon_{B,f}^{-1/4}E_{52}^{3/4}t_5^{-1/4}\nu^{-1/2}(1+z)^{-5/4}x^{-2} & \nu_{c,f}<\nu \le \nu_{m,f}\\
C_3\epsilon_{B,f}^{(p-2)/4}\epsilon_e^{(p-1)}E_{52}^{(p+2)/4}t_5^{-(3p-2)/4}\nu^{-p/2}(1+z)^{(p-6)/4}x^{-2} & \nu_{m,f} \le \nu \\
\end{array} \right.
\label{eq:reddepforfast}
\end{equation}
For the slow cooling case, we have
\begin{equation}
F_{\nu}=\left\{ \begin{array}{l@{\quad \quad}l}
 C_4\epsilon_{B,f}^{1/3}\epsilon_e^{-2/3}E_{52}^{5/6}t_5^{1/2}n^{1/2}\nu^{1/3}(1+z)^{-7/6}x^{-2} & \nu < \nu_{m,f} \\
C_5\epsilon_{B,f}^{(p+1)/4}\epsilon_e^{(p-1)}E_{52}^{(p+3)/4}t_5^{-3(p-1)/4}n^{1/2}\nu^{(1-p)/2}(1+z)^{(p-5)/4}x^{-2} & \nu_{m,f} \le \nu < \nu_{c,f} \\
 C_3\epsilon_{B,f}^{{(p-2)}/4}\epsilon_e^{(p-1)} E_{52}^{(p+2)/4}t_5^{-(3p-2)/4)}\nu^{-p/2}(1+z)^{(p-6)/4}x^{-2} & \nu_{c,f} \le \nu \\
 \end{array} \right.
 \label{eq:reddepforslow}
\end{equation}
where $C_1$=0.402, $C_2=8.65 \times 10^9$, $C_3=(8.6 \times 10^9)
\times (4.6 \times 10^{14})^{(p-1)/2}=8.5 \times 10^{20}\
(p=2.5)$, $C_4=0.071$,and $C_5=5.5 \times 10^3 \times (4.58 \times
10^{14})^{(p-1)/2}=5.4 \times 10^{14}\ (p=2.5)$ and $x$ is defined
as \beq x = [1-1.115(1+z)^{-1/2}]. \label{eq:x} \enq Here,
$\epsilon_{B,f}, \epsilon_e, E_{52}$ have been assumed to be constant
parameters, while others like $D_{28}(z)$ and $n$ may be redshift
dependent in different cases. $\nu$ is the observed
frequency. For simplicity, we present here the scaling relation
for the flux with several parameters which may change with
redshift. We substitute each of these quantities into the
expressions above, and get the scaling relations with redshift for
different cases.

Fast cooling case:
\begin{equation}
F_{\nu} \propto\left\{ \begin{array}{l@{\quad \quad}l}
n^{5/6}(1+z)^{-5/6}x^{-2} & \nu < \nu_{c,f} \\
n^0(1+z)^{-5/4}x^{-2} & \nu_{c,f} < \nu <\nu_{m,f}\\
n^0(1+z)^{(p-6)/4}x^{-2} & \nu_{m,f} < \nu \\
 \end{array} \right.
 \label{eq:scalforfast}
\end{equation}

Slow cooling case:

\begin{equation}
F_{\nu}\propto \left\{ \begin{array}{l@{\quad \quad}l}
n^{1/2}(1+z)^{-7/6}x^{-2} & \nu < \nu_{m,f} \\
n^{1/2}(1+z)^{(p-5)/4}x^{-2} & \nu_{m,f}<\nu<\nu_{c,f} \\
n^0(1+z)^{(p-6)/4}x^{-2} & \nu_{c,f}<\nu \\
 \end{array} \right.
 \label{eq:scalforslow}
\end{equation}

Substituting the redshift dependence of the number density into
the relations above gives straightforwardly the scaling relations
for the different density cases.

\subsection{Reverse Shock}
The scalings here are taken from Kobayashi (2000) and Zhang et al.
(2003). Here the crossing time is defined as $t_\times=max(T,
t_{dec})$, where $T$ is the burst duration and $t_{dec}$ is
defined as $[(3E/4\pi \eta^2 n m_p c^2)^{1/3}/2 \eta^2 c](1+z)$ in
the observer frame. Also, $\hat{\eta}={\rm min}(\eta,
\eta_c^2/\eta)$, where $\eta$ is the initial Lorentz factor and
$\eta_c$ is defined as the critical initial Lorentz factor
$\eta_c=125E_{52}^{1/8}n^{-1/8}T_2^{-3/8}(\frac{1+z}{2})^{3/8}$
(Zhang et al 2003). For the thin shell case, one has
$t_\times=t_{dec}$ and $\hat{\eta}=\eta$, while for the thick shell
case, one has $ t_\times = T$ and $\hat{\eta}=\eta_c^2/\eta$.

In the thick shell case, the typical parameters at crossing time
$t_\times$ are:

\begin{equation}
\nu_{m,r}(t_{\times})=(\hat{\eta})^{-2} {\cal R_B}
\nu_{m,f}(t_{\times}), \ \nu_{c,r}(t_{\times})={\cal R_B}^{-3}
\nu_{c,f}(t_{\times}), \ F_{\nu,max,r}(t_{\times})= \hat{\eta}
{\cal R_B} F_{\nu,max,f}(t_{\times})
 \label{eq:thickcrossing}
\end{equation}
where ${\cal R_B}\equiv B_r/B_f=
\left(\epsilon_{B,r}/\epsilon_{B,f}\right)^{1/2}$.

The scaling relations before and after the shock crossing time
$t_\times$ are

\begin{equation}
t<t_{\times}:~~~ \nu_{m,r} \propto t^0,\\\nu_{c,r} \propto t^{-1}, \\\
F_{\nu,m,r} \propto t^{1/2}
 \label{eq:thickbefore}
\end{equation}

\begin{equation}
t>t_{\times}:~~~ \nu_{m,r} \propto t^{-73/48} \simeq t^{-3/2},\\\nu_{c,r} \propto t^{-73/48} \simeq t^{-3/2}, \\\
F_{\nu,m,r} \propto t^{-47/48} \simeq t^{-1}
 \label{eq:thickafter}
\end{equation}

 For the thin shell case, at the crossing time, one has,

\begin{equation}
\nu_{m,r}(t_{\times})=\eta^{-2} {\cal R_B} \nu_{m,f}(t_{\times}),\
\nu_{c,r}(t_{\times})={\cal R_B}^{-3} \nu_{c,f}(t_{\times}), \
F_{\nu,m,r}(t_{\times})=\eta {\cal R_B} F_{\nu,m,f}(t_{\times})
 \label{eq:thincrossing}
\end{equation}

The scaling relations before and after the shock crossing time $t_\times$ are

\begin{equation}
t<t_{\times}:~~~ \nu_{m,r}=t^6,\\\nu_{c,r} \propto t^{-2}, \\\
F_{\nu,m,r} \propto t^{3/2}
 \label{eq:thinbefore}
\end{equation}

\begin{equation}
t>t_{\times}:~~~ \nu_{m,r} \propto t^{-54/35} \simeq t^{-3/2},\\\nu_{c,r} \propto t^{-54/35}\simeq t^{-3/2} , \\\
F_{\nu,m,r} \propto t^{-34/35} \simeq t^{-1}
 \label{eq:thinafter}
\end{equation}

Before the crossing time, for the thin shell case observed flux can be expressed as:

\begin{equation}
F_{\nu}=\left\{ \begin{array}{l@{\quad \quad}l}
[F_{\rm{\nu, f}} \nu_{\rm{c,f}}^{-1/3} \nu^{1/3}] [\hat{\eta} {\cal R_B}^2 (t/t_{\times})^{5/6}] & \nu<\nu_{c,r} \\
{[F_{\rm{\nu, f}} \nu_{\rm{c,f}}^{1/2} \nu^{-1/2}]} [\hat{\eta} {\cal R_B}^{-1/2} (t/t_{\times})^{1/2}] & \nu_{c,r}<\nu \le \nu_{m,r}\\
{[F_{\rm{\nu, f}} \nu_{\rm{m,f}}^{(p-1)/2}\nu_{\rm{c,f}}^{1/2} \nu^{-p/2}]} [ {\hat{\eta}}^{(2-p)}{\cal R_B}^{(p-2)/2} (t/t_{\times})^{(3p-5/2)}] & \nu_{m,r} \le \nu \\
\end{array} \right.
\label{eq:thinbefluxfast}
\end{equation}

\begin{equation}
F_{\nu}=\left\{ \begin{array}{l@{\quad \quad}l}
[F_{\rm{\nu, f}} \nu_{\rm{m,f}}^{-1/3} \nu^{1/3}] [{\hat{\eta}}^{(5/3)}{\cal R_B}^{2/3} (t/t_{\times})^{5/6}] & \nu<\nu_{m,r} \\
{[F_{\rm{\nu, f}} \nu_{\rm{m,f}}^{(p-1)/2} \nu^{-(p-1)/2}]} [{\hat{\eta}}^{(2-p)}{\cal R_B}^{(p+1)/2} (t/t_{\times})^{3p+3/2}] & \nu_{m,r}<\nu \le \nu_{c,r}\\
{[F_{\rm{\nu, f}} \nu_{\rm{m,f}}^{(p-1)/2}\nu_{\rm{c,f}}^{1/2} \nu^{-p/2}]} [ {\hat{\eta}}^{(2-p)}{\cal R_B}^{(p-2)/2} (t/t_{\times})^{(3p-5/2)}] & \nu_{c,r} \le \nu \\
\end{array} \right.
\label{eq:thinbefluxslow}
\end{equation}
while for the thick shell the flux is:

\begin{equation}
F_{\nu}=\left\{ \begin{array}{l@{\quad \quad}l}
[F_{\rm{\nu, f}} \nu_{\rm{c,f}}^{-1/3} \nu^{1/3}] [\hat{\eta} {\cal R_B}^2 (t/t_{\times})^{5/6}] & \nu<\nu_{c,r} \\
{[F_{\rm{\nu, f}} \nu_{\rm{c,f}}^{1/2} \nu^{-1/2}]}[ \hat{\eta}{\cal R_B}^{-1/2}]  & \nu_{c,r}<\nu \le \nu_{m,r}\\
{[F_{\rm{\nu, f}} \nu_{\rm{m,f}}^{(p-1)/2}\nu_{\rm{c,f}}^{1/2} \nu^{-p/2}]} [\hat{\eta}^{(2-p)}{\cal R_B}^{(p-2)/2}]  & \nu_{m,r} \le \nu \\
\end{array} \right.
\label{eq:thickbefluxfast}
\end{equation}

\begin{equation}
F_{\nu}=\left\{ \begin{array}{l@{\quad \quad}l}
[F_{\rm{\nu, f}} \nu_{\rm{m,f}}^{-1/3} \nu^{1/3}][\hat{\eta}^{(5/3)}{\cal R_B}^{2/3} (t/t_{\times})^{1/2}] & \nu<\nu_{m,r} \\
{[F_{\rm{\nu, f}} \nu_{\rm{m,f}}^{(p-1)/2} \nu^{-(p-1)/2}]} [\hat{\eta}^{(2-p)}{\cal R_B}^{(p+1)/2} (t/t_{\times})^{1/2}]  & \nu_{m,r}<\nu \le \nu_{c,r}\\
{[F_{\rm{\nu, f}} \nu_{\rm{m,f}}^{(p-1)/2}\nu_{\rm{c,f}}^{1/2} \nu^{-p/2}]}  [\hat{\eta}^{(2-p)}{\cal R_B}^{(p-2)/2}] & \nu_{c,r} \le \nu \\
\end{array} \right.
\label{eq:thickbefluxslow}
\end{equation}

After the crossing time, for both the thick shell and thin shell cases the expressions
for the flux are:

\begin{equation}
F_{\nu}=\left\{ \begin{array}{l@{\quad \quad}l}
[F_{\rm{\nu, f}} \nu_{\rm{c,f}}^{-1/3} \nu^{1/3}] [{\hat{\eta}}{\cal R_B}^2 (t/t_{\times})^{-1/2}] & \nu<\nu_{c,r} \\
0 & \nu_{c,r}<\nu \le \nu_{m,r}\\
0 & \nu_{m,r} \le \nu \\
\end{array} \right.
\label{eq:bothafterfast}
\end{equation}

\begin{equation}
F_{\nu}=\left\{ \begin{array}{l@{\quad \quad}l}
[F_{\rm{\nu, f}} \nu_{\rm{m,f}}^{-1/3} \nu^{1/3}] [{\hat{\eta}}^{5/3}{\cal R_B}^{3/2} (t/t_{\times})^{-1/2}] & \nu<\nu_{m,r} \\
{[F_{\rm{\nu, f}} \nu_{\rm{m,f}}^{(p-1)/2} \nu^{-(p-1)/2}]} [{\hat{\eta}}^{(2-p)}{\cal R_B}^{(p+1)/2} (t/t_{\times})^{(1-3p)/4}] & \nu_{m,r}<\nu \le \nu_{c,r}\\
0 & \nu_{m,r} \le \nu \\
\end{array} \right.
\label{eq:bothafterslow}
\end{equation}

Substituting the redshift dependence into equations above for the reverse shock,
we obtain similar scaling relations as those for the forward shock.

For the thin shell case before crossing time:

Fast cooling case:
\begin{equation}
F_{\nu} \propto\left\{ \begin{array}{l@{\quad \quad}l}
n^{19/18}(1+z)^{-5/6}x^{-2} & \nu < \nu_{c,r} \\
n^{13/12}(1+z)^{-5/4}x^{-2} & \nu_{c,r} < \nu <\nu_{m,r}\\
n^{(5p-4)/4}(1+z)^{(p-6)/4}x^{-2} & \nu_{m,r} < \nu \\
 \end{array} \right.
 \label{eq:scthinbefast}
\end{equation}

Slow cooling case:

\begin{equation}
F_{\nu}\propto \left\{ \begin{array}{l@{\quad \quad}l}
n^{11/18}(1+z)^{-7/6}x^{-2} & \nu < \nu_{m,r} \\
n^{(3p+3)/4}(1+z)^{(p-5)/4}x^{-2} & \nu_{m,r}<\nu<\nu_{c,r} \\
n^{(5p-4)/4}(1+z)^{(p-6)/4}x^{-2} & \nu_{c,r}<\nu \\
 \end{array} \right.
 \label{eq:scthinbeslow}
\end{equation}

For the thin shell case after the crossing time:

Fast cooling case:
\begin{equation}
F_{\nu} \propto\left\{ \begin{array}{l@{\quad \quad}l}
n^{11/18}(1+z)^{-5/6}x^{-2} & \nu < \nu_{c,r} \\
0 & \nu_{c,r} < \nu <\nu_{m,r}\\
0 & \nu_{m,r} < \nu \\
 \end{array} \right.
 \label{eq:scthinaffast}
\end{equation}

Slow cooling case:

\begin{equation}
F_{\nu}\propto \left\{ \begin{array}{l@{\quad \quad}l}
n^{1/6}(1+z)^{-7/6}x^{-2} & \nu < \nu_{m,r} \\
n^{1/3}(1+z)^{(p-5)/4}x^{-2} & \nu_{m,r}<\nu<\nu_{c,r} \\
0 & \nu_{c,r}<\nu \\
 \end{array} \right.
 \label{eq:scthinafslow}
\end{equation}

One sees that if the number density around the GRBs does not
change with the redshift, i.e. $n=$ constant for all $z$, the
reverse shock emission depends on redshift in the same way as the
forward shock emission. However, in the $n \propto (1+z)^4$ case,
the behavior for the reverse and forward shock emission is
different. The thick shell behavior can be obtained in a similar
way.

 \end{document}